# Upward band gap bowing and negative mixing enthalpy in multi-component cubic halide perovskite alloys

Xiuwen Zhang[1,†], Fernando P. Sabino[2], Jia-Xin Xiong[1], and Alex Zunger[1,*]

[1] Renewable and Sustainable Energy Institute, University of Colorado, Boulder, Colorado 80309, USA

[2] Department of Materials Engineering, São Carlos School of Engineering, University of São Paulo, São Carlos, SP 13563-120, Brazil

Physical properties *intermediate* between constituents of alloys can be achieved as downward convex positive bowing, upward concave negative bowing, or zero bowing. Such bowing effects are essential for band gap engineering in semiconductor alloys. Upward band gap bowing effects are rather rare, hindering the exploration on half of the available physical property space of alloys. Part of the this being a rare event is related to the need to stabilize an alloy with low mixing enthalpy, so it does not phase separate. In this paper we find via density functional theory that one can satisfy the simultaneous conditions of negative mixing enthalpy and upward band gap bowing in four-component $ABX_3$ halide perovskite alloys in the cubic perovskite structure. Such perovskite alloys have the B-site occupied by a mixture of group IVB and IIB elements that have the IVB-*s* and IIB-*s* states in the valence bands and conduction bands, respectively, leaving the delocalized *s* states strongly repel each other. This *s*-*s* repulsion leads to the upward band gap bowing and negative mixing enthalpies simultaneously. Remarkably, we identify a perovskite alloy that has a band gap much larger than all its components. Analogous trends of upward band gap bowing and negative mixing enthalpy also appear in the corresponding three-component and two-component $ABX_3$ halide perovskite alloys. These observations of upward band gap bowing and negative mixing enthalpy will significantly accelerate the design of stable upward band gap bowing alloys in a broad range of material families.

[†] Corresponding author: xiuwen.zhang@colorado.edu

[*] Corresponding author: alex.zunger@colorado.edu

# I. INTRODUCTION

Alloys are often made in composition x to achieve properties P(x) intermediate between those of the constituents. Such *intermediate* behavior can take the form of *downward* bowing (positive bowing coefficient $b > 0$) or upward bowing (negative bowing coefficient $b < 0$) or no bowing ($b \sim 0$), as shown in Fig. 1(a). Although the bowing of the band gap in semiconductors is a critical effect for electronic and optical engineering of semiconductors, most bowing effects reported in experiment [1-8] and theory [9-15] are downward bowing. Typical material platforms having the downward band gap bowing effect are isovalent zinc-blende, isovalent binary perovskites, isovalent chalcopyrites, etc. [1-15]. So far, the upward band gap bowing effect was only observed in binary hetero-valent bulk alloy $CsCd_xPb_{1-x}Br_3$ [16], and predicted in binary isovalent epitaxial alloy InAs/GaAs [17] and ternary hetero-valent bulk alloy $Cs(Pb_xSn_yCd_{1-x-y})Br_3$ [18]. The series of hetero-valent bulk alloys $CsCd_xPb_{1-x}Br_3$ [16] and $Cs(Pb_xSn_yCd_{1-x-y})Br_3$ [18] raises the interest to design upward band gap bowing effects in halide perovskite materials. However, the material space for binary and ternary halide perovskite alloys with the desired electronic properties for realizing upward band gap bowing is very limited. Furthermore, the mechanisms for upward band gap bowing in the hetero-valent bulk halide perovskite alloys are currently not well understood, let alone the relationship between the material functionality upward band gap bowing and material stability. The dual conditions of material functionality and stability are often difficult to achieve due to their contraindicative relationship [19]. Seeking electronic configurations that can enable the target material functionality and stability simultaneously can significantly accelerate the design of functional materials for technological applications. Halide perovskite alloys in the cubic perovskite structure [space group (SG): Pm-3m] can be made with multi-components, exemplified by the fluoride perovskite alloy with 1:1:1:1:1 composition $K(MgMnFeCoNi)F_3$ as a single-phase alloy without phase separation experimentally [20]. The stability of multi-component alloys with equal quantities of components are found to benefit from the high entropy of the systems [21,22]. These multi-component alloys offer the opportunity to design more halide perovskite systems with stronger upward band gap bowing and in the meantime being stable and even lead free.

In this paper we find via density functional theory (DFT) that one can satisfy the simultaneous conditions of low mixing enthalpy or formation energy (i.e. not readily phase separated) and upward band gap bowing. The DFT details are introduced in Methods section and Appendix A. Upward band gap bowing effects not only offer larger band gaps and optical blueshift in semiconductor alloys but also introduce a different type of semiconductor interface with a wide-gap barrier layer or tunneling layer formed by the alloy due to the interface atomic interchange between the two narrow-gap semiconductors forming the interface. Interestingly, the enabling conditions involve (i) multi components and (ii) B-cations coming from different columns of the Periodic Table, such as IVB (e.g., Ge, Sn, Pb) and IIB (e.g., Cd). Examples of such combinations include quaternary cubic perovskite alloys $Cs_4[GeSnPbCd]I_{12}$ (with coherent mixing enthalpy $\Delta H_{coh.}$ = -6.3 meV/atom and excess band gap relative to components $\Delta E_g$ = 0.79 eV), $Cs_4[GePbCaCd]I_{12}$ (-6.2 meV/atom, 0.388 eV), and $Cs_4[GeSnPbCd]Br_{12}$ (-1.9 meV/atom, 0.48 eV) at 1:1:1:1 composition, possessing negative $\Delta H_{coh.}$. We reveal the mechanisms for the predicted upward band gap bowings based on the electronic configurations in the alloys. We further demonstrate the intriguing relationship between the band bowing effects and materials stabilization in the calculated multi-component alloys. Interestingly, analogous trends also appear in ternary and binary cubic perovskite alloys. Furthermore, many of the best candidate multi-component halide perovskite alloys predicted in this work are lead free. Since the mixing enthalpies for the two-component and three-component B-site halide perovskite alloys are found to be theoretically low [18], it is thermodynamically possible to pack multiple perovskite components in the alloy. It is interesting to explore the large material space of these multi-component alloys. Furthermore, multi-component alloys can generate multiple interesting properties and is easier explored theoretically first than via synthesis and characterization. Here, we focus on the multi-component halide perovskites alloys $Cs_4[BB'B''B''']X_{12}$ with X = Br or I, and B, B', B'', and B''' are chosen from column IVB (Ge, Sn, Pb), column IIB (Cd), or column IIA (Ca, Sr, Ba) (for comparison with IIB). Indeed, we find there are much more negative band gap bowing materials among the multi-component perovskites alloys than two-component perovskites alloys and the multi-component perovskites alloys produces a stronger band gap bowing effect. Our study suggests a group of stable halide perovskite alloys with large upward band gap bowing effects for optoelectronic and new energy applications and proposed a functional material design strategy that naturally fulfils desired material functionality and stability simultaneously.

**(a) Band gap variation $\Delta E_g(x)$ of cubic-shaped perovskite alloys**

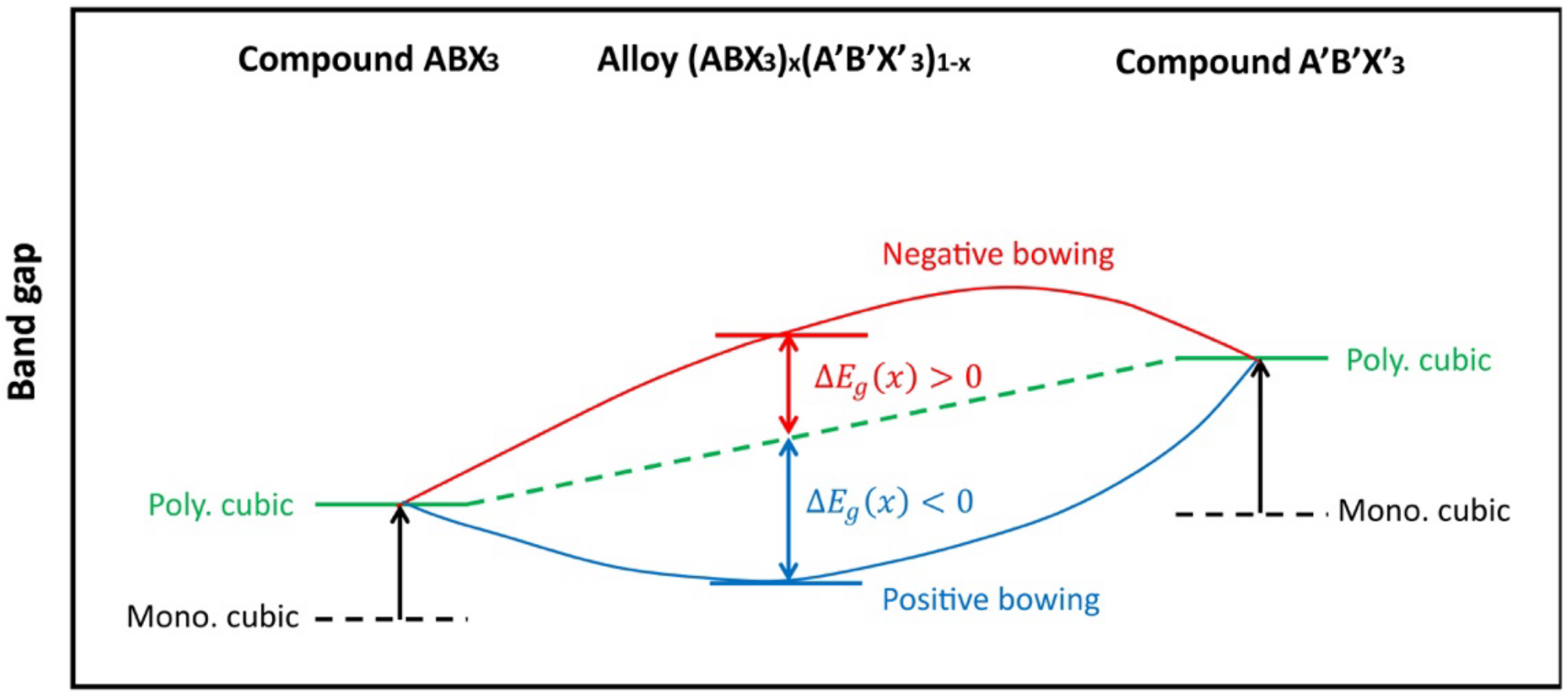

**(b) Coherent mixing enthalpies $\Delta H_{coh.}(x)$ of cubic-shaped perovskite alloys**

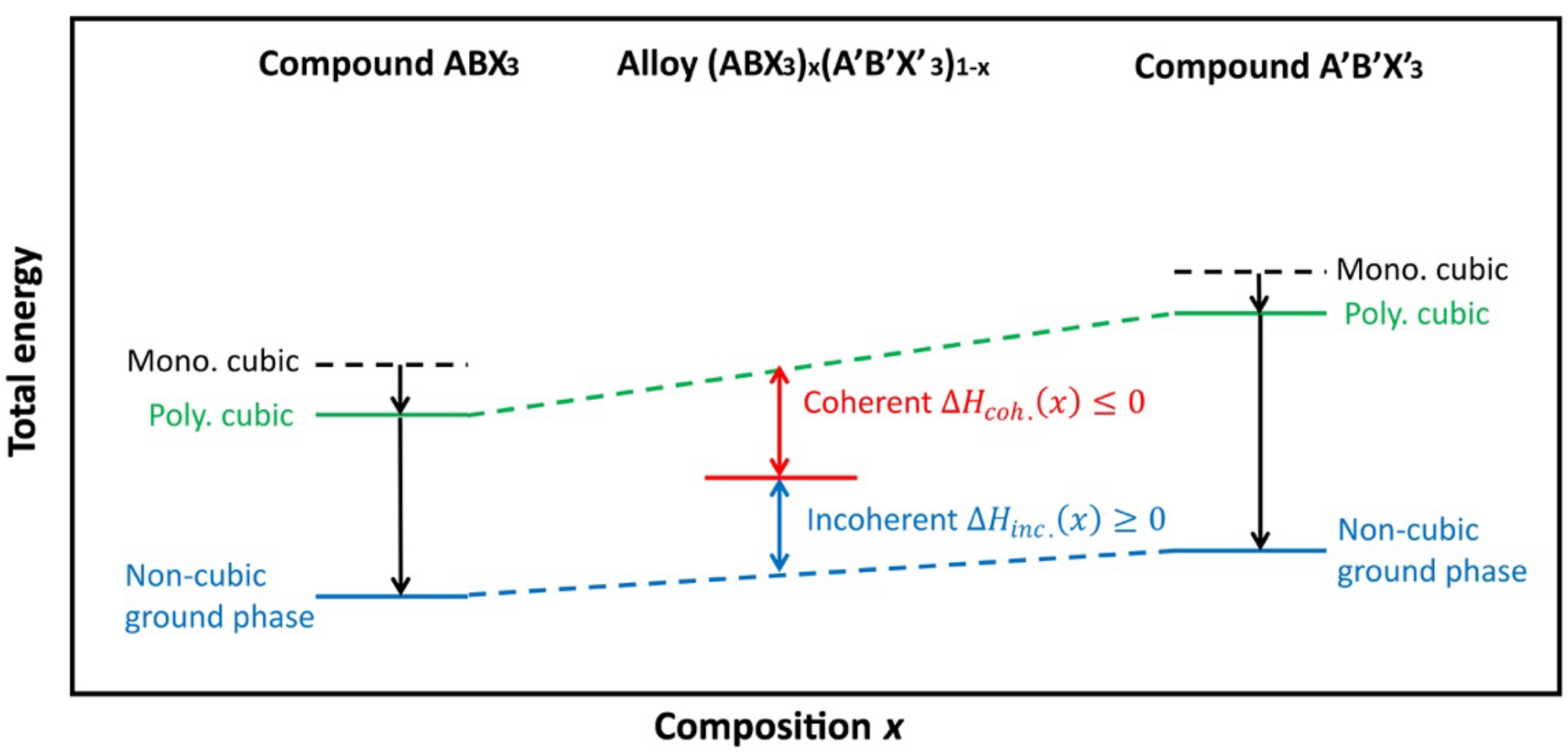

FIG. 1. Illustration of band gap bowing (a) as well as coherent and incoherent mixing enthalpies (b) in cubic perovskite alloys.

## II. METHODS

***Monomorphous and polymorphous materials:*** It is found that pure non-alloyed Pm-3m perovskite (and other) compounds can lower the energy by creating a distribution of local motifs of the same chemical identity such as $BX_6$ [23-25]. On the other hand, the energy lowering in non-cubic single $ABX_3$ phases (e.g. Pnma non-perovskite structure of $CsPbBr_3$ [26]) is smaller because, in part, the unit cell already includes different chemically identical motifs. We use a $2\sqrt{2} \times 2\sqrt{2} \times 4$ supercell of the cubic perovskite (Pm-3m) structure that can describe the polymorphous system [23-25], with quasi-randomly distributed four types of atoms on the B sites generated based on the special quasi-random structures (SQSs) method [27] by using the Alloy

Theoretic Automated Toolkit (ATAT) [28], followed by a random displacement of the atoms and structural relaxation. We fix the shape of the supercell to be $2\sqrt{2} \times 2\sqrt{2} \times 4$ to better describe the polymorphous system [23-25]. For the $2\sqrt{2} \times 2\sqrt{2} \times 4$ supercell, we used the best candidate with the better convergence of the correlation between second and third neighbors.

We used the same type of supercell structure and parameters for calculating the polymorphous $ABX_3$ structures as the polymorphous alloys. We also calculated the total energies of the $ABX_3$ for the seven stable or metastable ordered structures as observed in experiments for the $ABX_3$ compounds with A = Cs, and X = Br or I, and B from group IVB (Ge, Sn, Pb), group IIA (Ca, Sr, Ba), or group IIB (Cd) [26,29-34], sort out the lowest-energy structure for each $ABX_3$ compound, and compare their total energies with the polymorphous perovskite structures (Fig. 2) (these $ABX_3$ compounds as the alloy precursors in this work, including $CsPbI_3$ [33,35], $Cs(Ba,Sr)(Br,I)_3$ [31,36], $CsCdI_3$ [37,38], $CsCaI_3$ [31], $CsPbBr_3$ [26], $CsSnI_3$ [29,34], $CsSnBr_3$ [39], $CsCaBr_3$ [31], $CdCdBr_3$ [30], and $CsGe(I,Br)_3$ [32,40], have been realized in experiments) [41]. For certain $ABX_3$ compounds, the polymorphous perovskite structure has nearly the same total energy as the lowest-energy structure (Fig. 2).

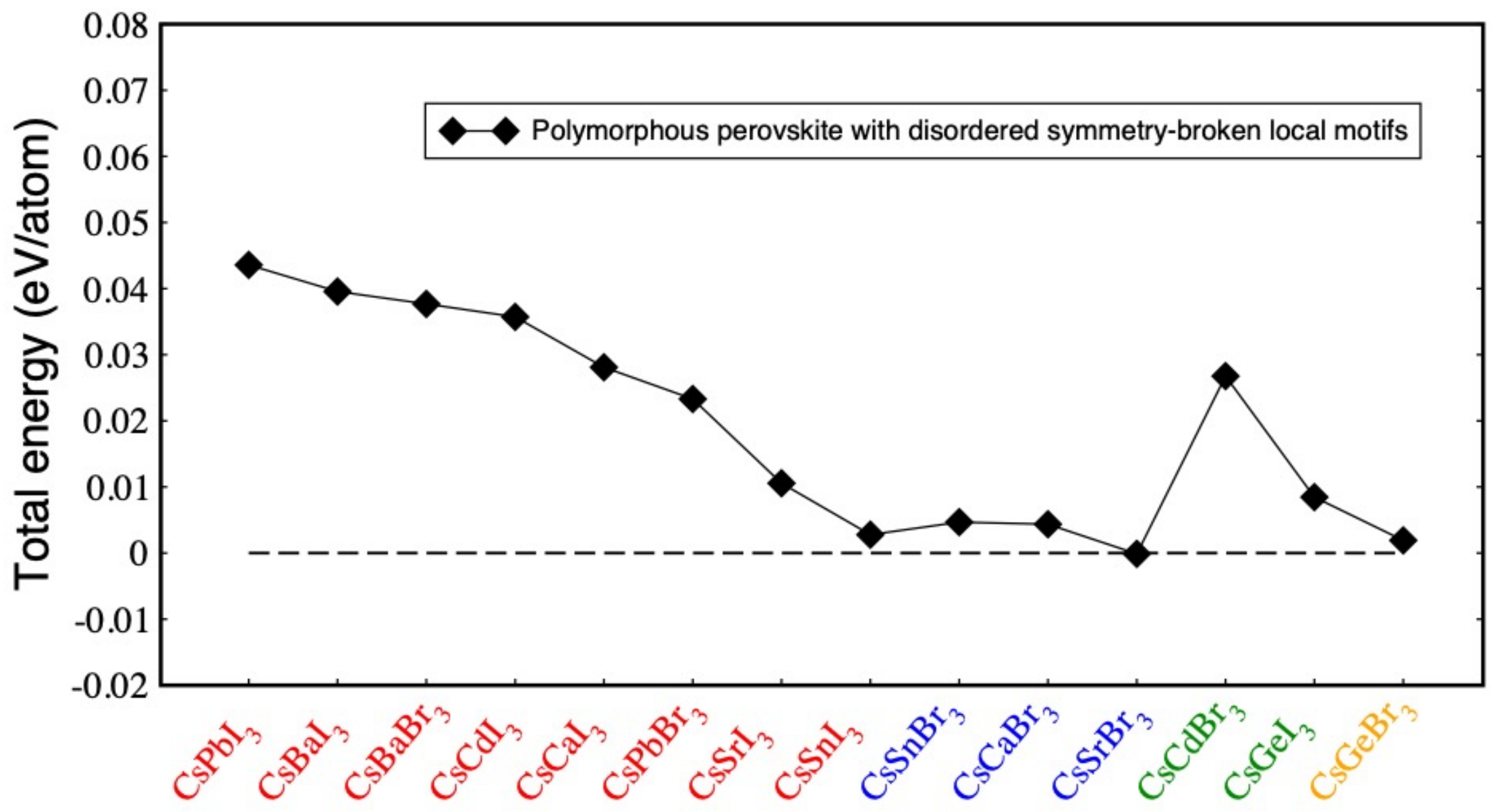

FIG. 2. Total energy of the polymorphous structure of the $ABX_3$ compounds relative to the lowest-energy structure of the seven stable or metastable ordered structures as observed in experiments for the $ABX_3$ with A = Cs, X = Br or I, and B = Ge, Sn, Pb, Ca, Sr, Ba, or Cd [26,29-34]. Red font: compounds with the orthorhombic $CsPbBr_3$-type lowest-energy structure (SG: Pnma). Blue font: compounds with the orthorhombic perovskite structure (SG: Pnma). Green font: compounds with the hexagonal $CsCdBr_3$-type structure (SG: $P6_3/mmc$). Orange font: compounds with the rhombohedral $CsGeBr_3$-type structure (SG: R3m).

***Excess band gap and band gap bowing:*** Excess band gap of a polymorphous alloy relative to the linear interpolation of band gaps of pure compounds is calculated as [see Fig. 1(a)]:

$$\Delta E_g = E_g^{alloy} - \sum_{i=1}^{N} E_g^i x^i, \quad (1)$$

where $E_g^{alloy}$ is the band gap of the polymorphous alloy, $E_g^i$ is the band gap of the *i*th component calculated using the same type of supercell structure as the polymorphous alloy, $x^i$ is the composition of the *i*th component, and *N* is the number of components. The band gap bowing coefficient *b* is calculated by:

$$b = -\frac{\Delta E_g}{\prod_i x^i}. \quad (2)$$

Note that the excess band gap $\Delta E_g > 0$ means upward band gap bowing, or negative band gap bowing, which is reduced in binary alloys to Eq. (18) in Ref. [42] but with a negative sign on the band gap bowing coefficient.

***Coherent and incoherent mixing enthalpies:*** The coherent mixing enthalpy of cubic polymorphous alloy [see Fig. 1(b)] is calculated with respect to the linear interpolation of the total energies of the components in the same type of polymorphous cubic structure as the polymorphous alloy:

$$\Delta H_{coh.} = E^{alloy} - \sum_{i=1}^{N} E_{p.c.}^i x^i \quad (3)$$

where $E^{alloy}$ is the total energy of the alloy per atom, and $E_{p.c.}^i$ is the total energy of the polymorphous cubic structure of the *i*th component per atom. The coherent mixing enthalpy $\Delta H_{coh.}$ can either be positive or negative. In this work, we focus on the coherent mixing enthalpy $\Delta H_{coh.}$, especially the negative $\Delta H_{coh.}$. As and aside, there is another mixing enthalpy, the incoherent mixing enthalpy, which is calculated with respect to the linear interpolation of the total energies of the components per atom in the lowest-energy structures (being non-cubic) in the physical equilibrium:

$$\Delta H_{inc.} = E^{alloy} - \sum_{i=1}^{N} E_{n.c.}^i x^i \quad (4)$$

where $E_{n.c.}^i$ is the total energy of the lowest-energy structures (being non-cubic) of the *i*th component. The incoherent mixing enthalpy $\Delta H_{inc.}$ is always positive.

# III. RESULTS

## III.A Upward band gap bowing: DFT results

As shown in Tables I and II in Appendix B, we find 14 four-component halide perovskite alloys from the $Cs_4[BB'B''B''']X_{12}$ family that demonstrate upward band gap bowing according to HSE+SOC calculations. Examples include $Cs_4[GeSnPbCd]I_{12}$ ($\Delta E_g^{alloy}$ = 0.79 eV) (Fig. 3) and $Cs_4[GeSnPbCd]Br_{12}$ ($\Delta E_g^{alloy}$ = 0.48 eV). As comparisons, analogous compounds $Cs_4[GeSnPbCa]I_{12}$ ($\Delta E_g^{alloy}$ = -0.23 eV) and $Cs_4[GeSnPbCa]Br_{12}$ ($\Delta E_g^{alloy}$ = -0.43 eV) have downward band gap bowing. In Tables I and II in Appendix B, we separate the four-component alloys into seven chemical groups depending on the identity of the B-site elements. It is interesting to see that only the chemical groups with IIB B-site elements can have upward band gap bowing. Among these IIB-containing chemical groups in iodide systems, only the chemical groups with three (IVB-IVB-IVB-IIB) or two (IVB-IVB-IIA-IIB) IVB elements have host upward band gap bowing effects. For bromide systems, the IVB-IVB-IIA-IIB chemical group is separated into Ge-Pb-IIA-IIB subgroup that host upward band gap bowing and the Ge-Sn-IIA-IIB and Sn-Pb-IIA-IIB subgroups that lack the upward band gap bowing. The difference between the Ge-Pb-IIA-IIB versus the Ge-Sn-IIA-IIB and Sn-Pb-IIA-IIB subgroups could be related to the volume effect (Ge mixed with Pb inducing large volume variation) and the spin-orbit coupling (SOC) effect (Pb inducing strong SOC effect) on the upward band bowing [18]. It is highly unusual to find that the halide perovskite alloy $Cs_4[GeSnPbCd]I_{12}$ has a band gap (1.96 eV) that is much larger than all its $ABX_3$ components (1.01 eV to 1.53 eV) [see Fig. 3(a)]. From Fig. 3(a), we can see that the upward band gap bowing of $Cs_4[GeSnPbCd]I_{12}$ relative to its components is very strong, corresponding to a band gap bowing coefficient of $b$ = -202.9 eV [Eq. (2)].

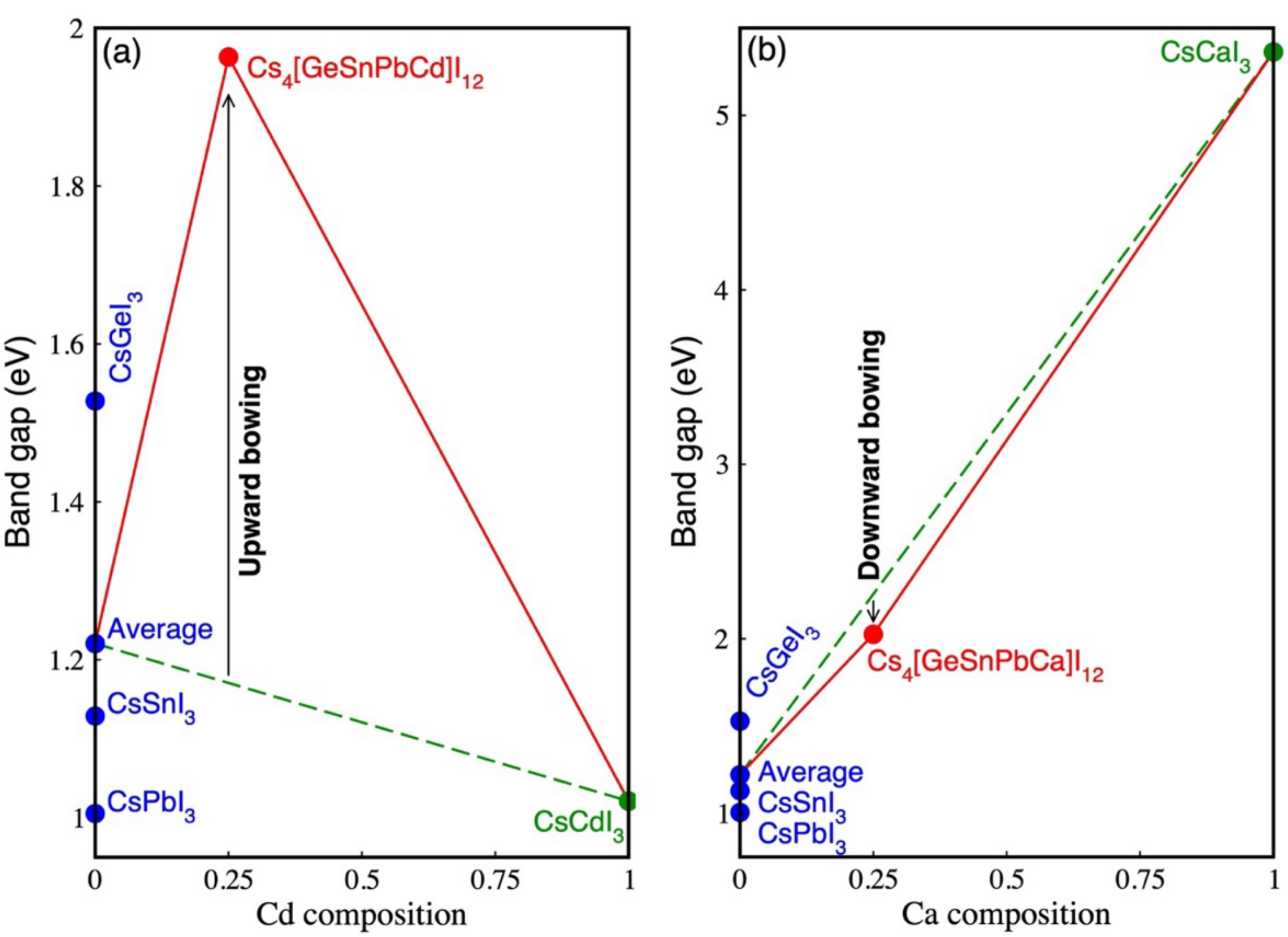

FIG. 3. Band gap bowing in $Cs_4[GeSnPbCd]I_{12}$ and $Cs_4[GeSnPbCa]I_{12}$. (a) Band gap of $Cs_4[GeSnPbCd]I_{12}$ relative to the components. (b) Band gap of $Cs_4[GeSnPbCa]I_{12}$ relative to the components.

## III.B Correlation between band gap bowing and coherent mixing enthalpy

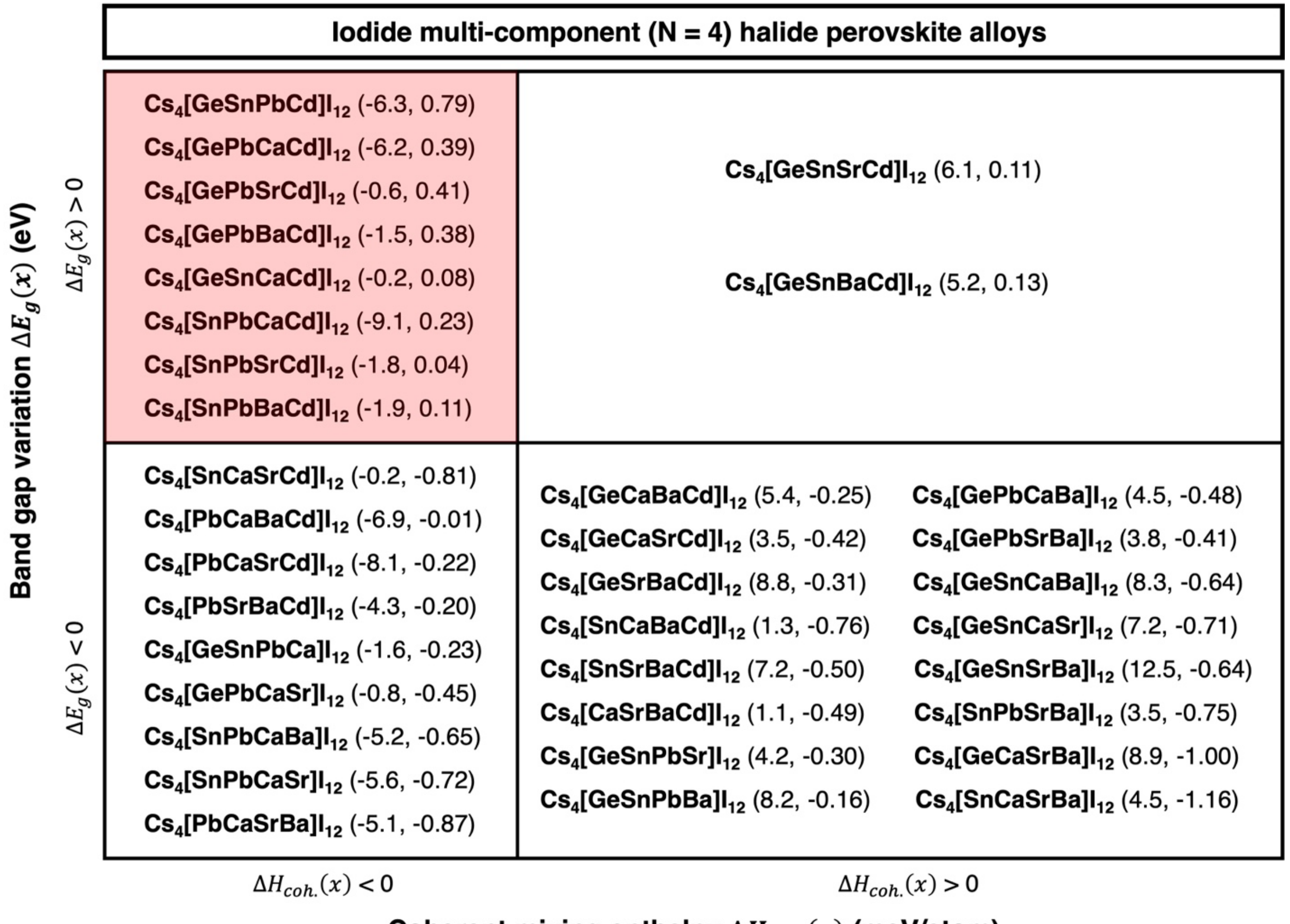

FIG. 4. Four categories of **iodide** N=4 cubic halide perovskite alloys at composition $x$=0.25 based on whether excess band gap $\Delta E_g(x)$ and the coherent mixing enthalpy $\Delta H_{coh.}(x)$ is negative or positive.

As shown in Tables I and II in Appendix B, surprisingly, we find that quite a few of the four-component cubic halide perovskite alloys $A_4[BB'B''B''']X_{12}$ have negative coherent mixing enthalpy $\Delta H_{coh.}$, [Eq. (3)] relative to the polymorphous cubic single perovskite constituents. The latter are calculated for consistency using the same type of supercell structure for the polymorphous alloy structures. This is due to the ease of creating interpenetrating networks in the perovskite framework. We further calculated the *incoherent* mixing enthalpy $\Delta H_{inc.}$,[ Eq. (4)] of the four-component halide perovskite alloys $A_4[BB'B''B''']X_{12}$ with respect to the lowest-energy structures of the $ABX_3$ constituents. As expected, we find that all incoherent alloys have positive formation enthalpies (see Tables I and II in Appendix B). Here, we mainly focus on the coherent mixing enthalpy, along with the upward band gap bowing magnitude $\Delta E_g$ [Eq. (1)] of the alloys.

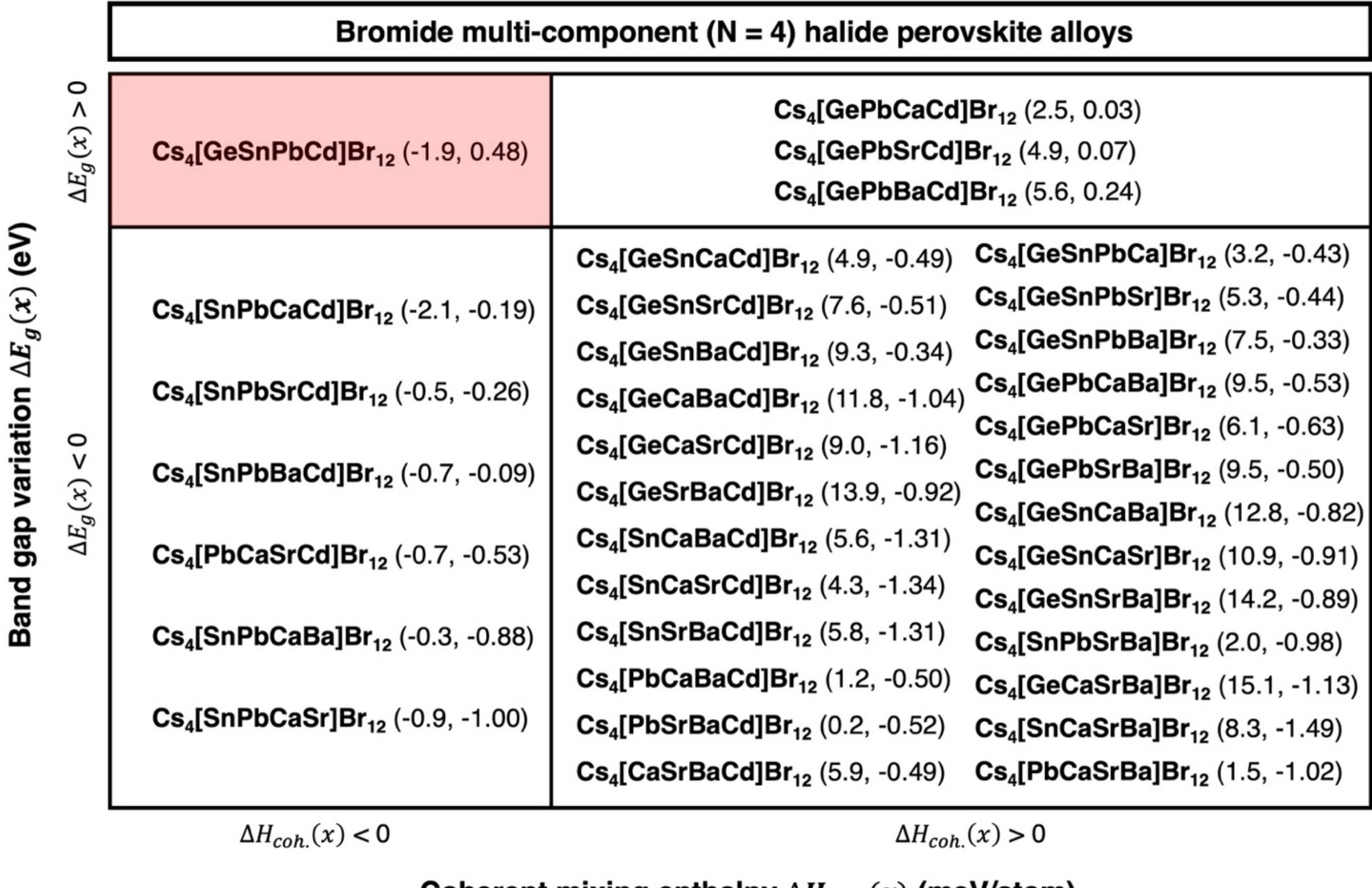

FIG. 5. Four categories of **bromide** N=4 cubic halide perovskite alloys at composition $x$=0.25 based on whether excess band gap $\Delta E_g(x)$ and coherent mixing enthalpy $\Delta H_{coh.}(x)$ is negative or positive.

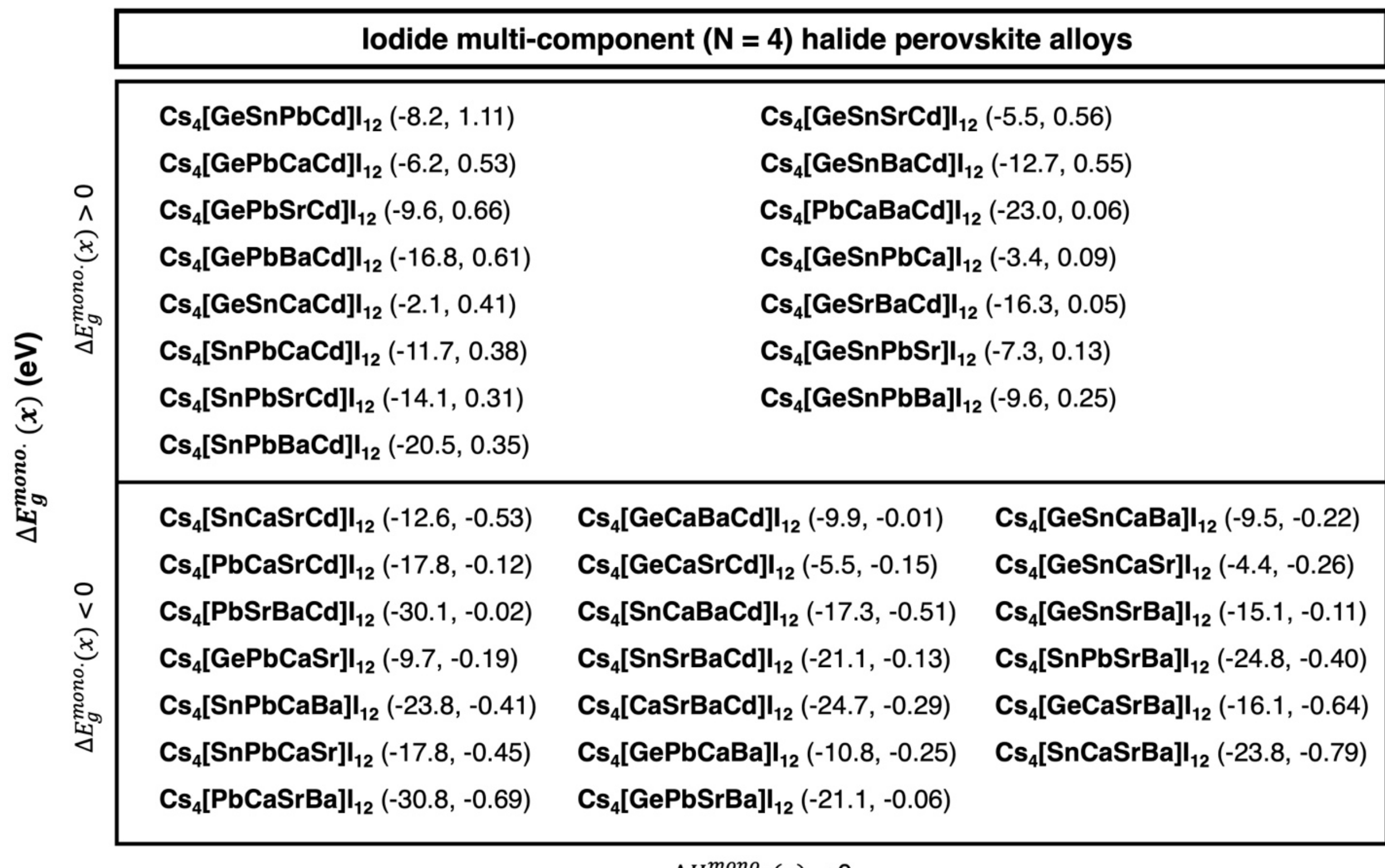

FIG. 6. Two categories of **iodide** N=4 cubic halide perovskite alloys at composition $x$=0.25 based on whether excess band gap $\Delta E_g^{mono.}(x)$ relative to monomorphous cubic $ABX_3$ is negative or positive. All the iodide alloys calculated have negative (coherent) mixing enthalpy $\Delta H_{mono.}(x)$ relative to monomorphous cubic $ABX_3$.

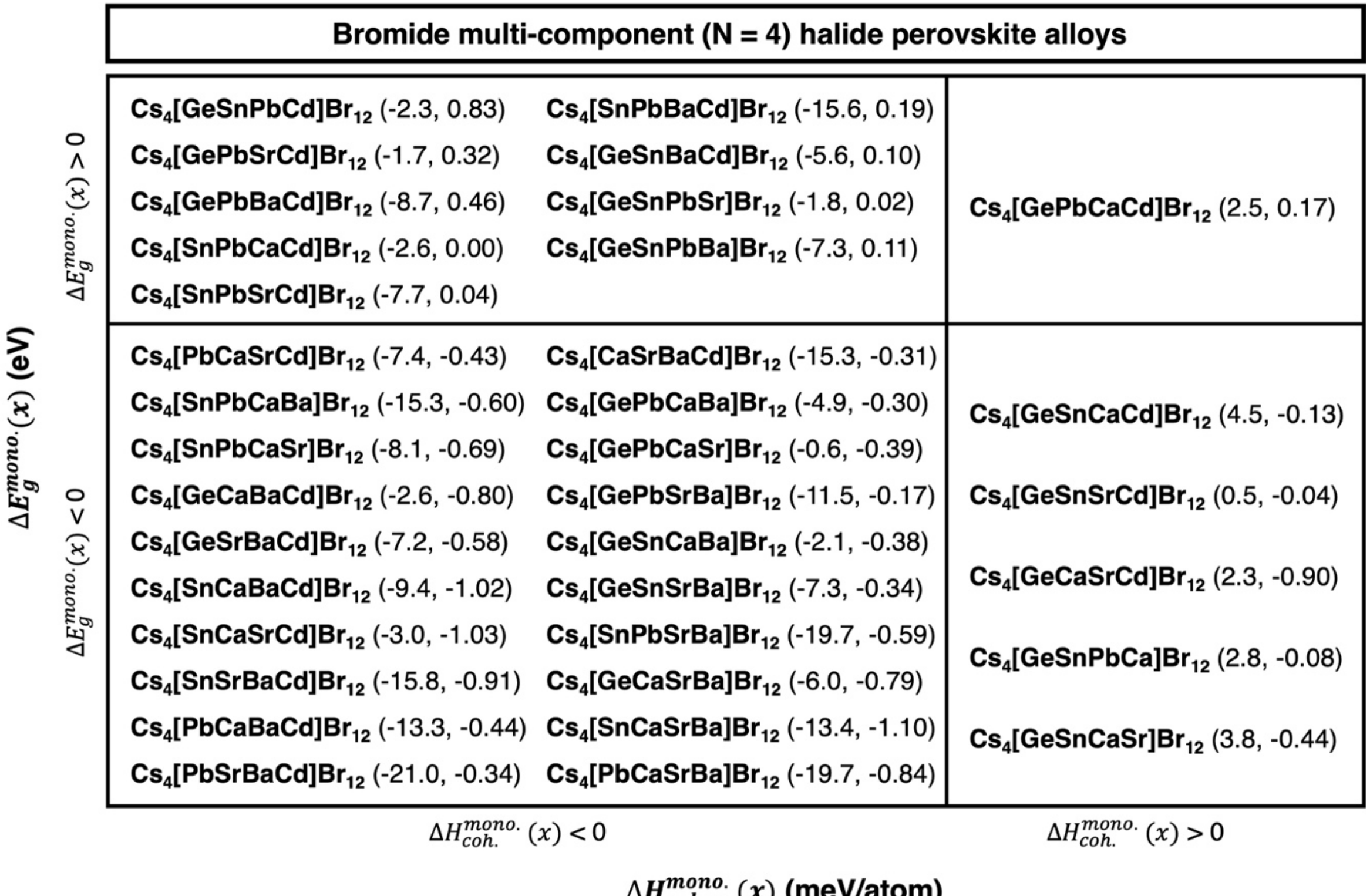

FIG. 7. Four categories of **bromide** N=4 cubic halide perovskite alloys at composition $x$=0.25 based on whether excess band gap $\Delta E_g^{mono.}(x)$ and (coherent) mixing enthalpy $\Delta H_{mono.}(x)$ relative to monomorphous cubic $ABX_3$ is negative or positive.

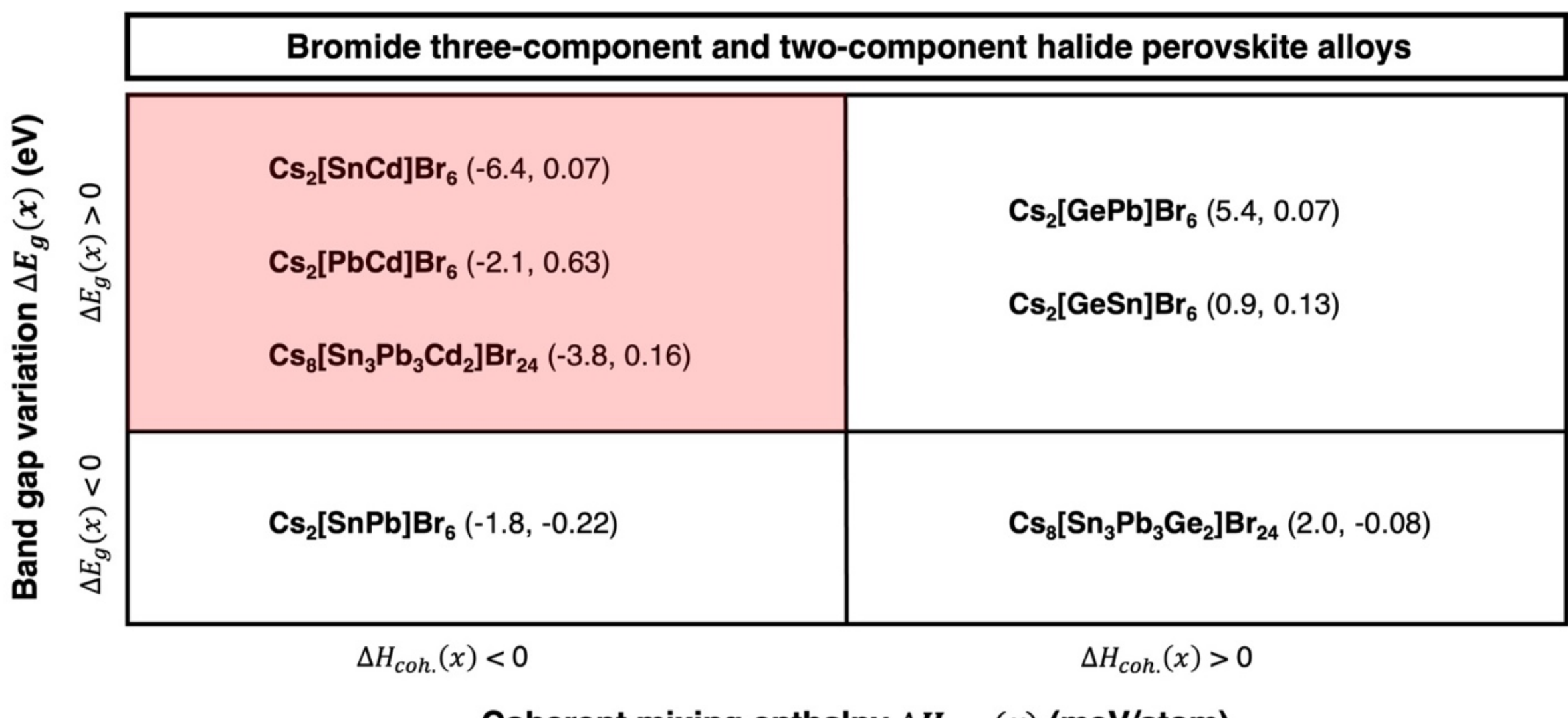

FIG. 8. Four categories of bromide N=2 and N=3 cubic halide perovskite alloys based on whether excess band gap $\Delta E_g(x)$ and coherent mixing enthalpy $\Delta H_{coh.}(x)$ is negative or positive. The data are taken from Ref. [18].

Band gap bowing effects and coherent mixing enthalpies are usually thought to be in different realms [9,11-16,18]. However, we find from Figs. 4 and 5 that there is a trend that four-component halide perovskite alloys $A_4[BB'B''B''']X_{12}$ with negative (positive) band gap bowing tends to have negative (positive) coherent mixing enthalpies. The Cd-containing alloys if with enough amount of group IVB elements tends to have upward band gap bowing and negative coherent mixing enthalpies (or nearly negative coherent mixing enthalpies) simultaneously.

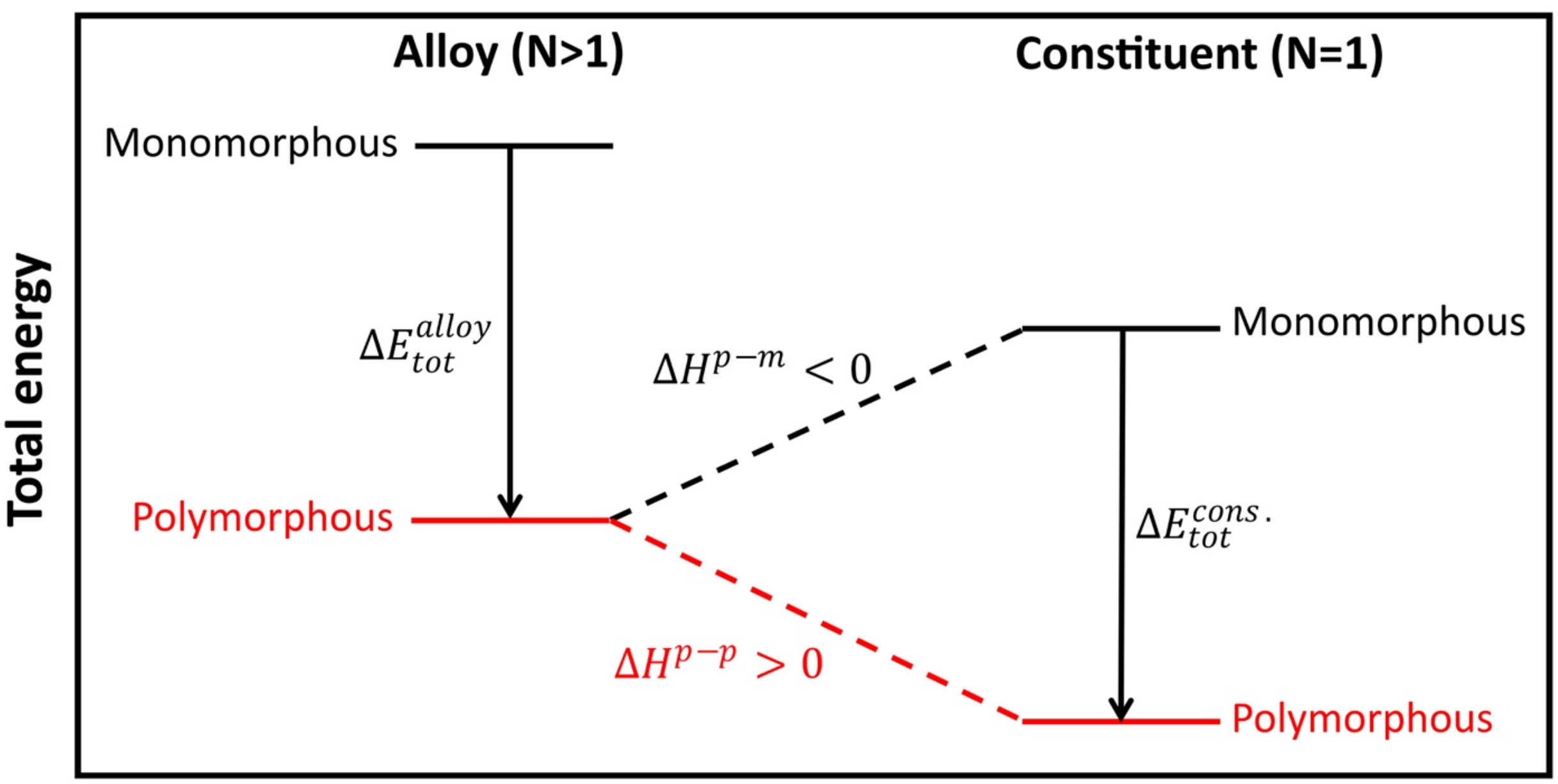

FIG. 9. Schematic diagram showing the relationship between total energies of polymorphous versus monomorphous halide perovskite alloys and alloy constituents.

To assess how important is the possible polymorphous nature of the materials, we compare the results with respect to individual perovskite constituents that are polymorphous (many different local environments) and the results with respect to monomorphous cubic perovskites (ideally ordered individual constituents having a single local environment). As Tables I and II in Appendix B list the coherent mixing enthalpies and the excess band gaps of the four-component halide perovskite alloys taken relative to the polymorphous constituents and Figs. 4 and 5 represent these results graphically, Tables III and IV in Appendix B list the same quantities of the alloys but relative to the monomorphous cubic constituents $ABX_3$ and Figs. 6 and 7 depict these results graphically.

*Polymorphous relaxation for the individual $ABX_3$ perovskites constituents*: There are generally well-known symmetry-controlled cell-internal degrees of freedom that need to be relaxed to minimum energy of the respective phase. For the orthorhombic phase (Pnma) for example, the cell internal degrees of freedom consist of octahedral tilting, off-center atomic shifting, etc. For cubic symmetry (Pm-3m), there is no cell internal degrees of freedom. The relevant energy lowering is indicated in Fig. 9 by the vertical error connecting the idealized monomorphous (*m*) structure with the polymorphous ($p$) structure where chemically identical $BX_6$ polyhedral can have different, energy lowering local environments. This subject of polymorphous energy lowering in individual constituents $ABX_3$ is well studied previously (Refs. [23-25]). We see from the results using monomorphous constituents (ideally ordered and with a single local environment) that the alloy stability and the upward band gap bowing are falsely overestimated as compared to the results when using the lower energy polymorphous cubic constituents. For example, the calculated iodide alloys have falsely negative mixing enthalpies relative to the monomorphous cubic constituents $ABX_3$ because the monomorphous constituent have abnormally high-energy, not benefitting from polymorphous relaxation, as illustrated schematically in Fig. 9.

*Polymorphous relaxation of the multi-component alloys:* We find that the alloy structures always have many different local environments, being polymorphous with local symmetry breaking. Experimentally, local lattice distortions that break local symmetry have been observed in different types of alloys [43,44]. The importance of symmetry breaking in halide perovskites was demonstrated in Refs. [25,45]: It significantly affects the physical properties including band gaps and stabilities. We did not consider the artificial case of monomorphous alloys $A_4[BB'B''B''']X_{12}$ with a single geometrical local environment, since polymorphous relaxation will significantly reduce the total energy in alloys. This can be illustrated in Fig. 9: Here one relaxes all the individual octahedra in a supercell leading to an energy-lowering alloy polymorphous network studied theoretically in Refs. [23-25].

*Relevant points to the study of alloy energetics*: Figure. 9 illustrates a number of: (i) Ignoring polymorphous relaxation of the constituents and using instead the monomorphous energy of the constituents to estimate the alloy formation energy *p-m* can easily lead to false negative formation energy on account of the too high monomorphous energy (see e.g. Figs. 6-7). (ii) The polymorphous relaxation in the alloy depends on the chemical composition and size mismatch but is generally thought to be larger than the polymorphous relaxation of the concentration-weighted constituents. The alloy formation energy *p-p* is generally positive but far less positive than for monomorphous alloy. We see that the whole discussion of multi-component alloy energetics depends on a balanced consideration of polymorphous relaxation of both constituents and the alloy.

It is interesting to note that similar trends to the upward band gap bowing, and negative coherent mixing enthalpy also appear in the corresponding three-component and two-component cubic halide perovskite alloys that have Cd, as shown in Fig. 8. We understand that both upward band gap bowing and coherent mixing enthalpies have complicated contributing factors. Our observation of the similar trends in Figs. 4, 5, and 8 hints that there is an interesting mechanism that can correlate upward band gap bowing with coherent mixing enthalpies.

## III.C The mechanism for upward band gap bowing and its correlation with negative coherent mixing enthalpy

To understand the above noted trends of upward band gap bowing in the $A_4[BB'B''B''']X_{12}$ halide perovskite alloys, we analyze the interaction of atomic orbitals in the alloys. Tables V and VI in Appendix C show the atomic orbital components of the band-edge states at the conduction band minimum (CBM) or the valence band maximum (VBM) of the $A_4$[BB′B′′B′′′]$X_{12}$ halide perovskite alloys. In the IVB-IVB-IVB-IIB and IVB-IVB-IIA-IIB groups of $A_4$[BB′B′′B′′′]$X_{12}$ alloys that can host upward band gap bowing (see Tables I and II in Appendix B), B-site atoms with $s^2p^2$ configuration (e.g. Sn and Pb) will have the $s^2$ orbital occupied in the valence band, and B-site atoms with $s^2p^0$ configuration (e.g. Cd and Sr) will have the $s^2$ orbital in the conduction band. Especially, when the atoms with $s^2p^0$ shell are from group IIB in the periodic table, here Cd, the *s* states of Cd are found to be a significant component (see Tables V and VI in Appendix C) at the CBM, which could offer the opportunity for the *s* states of IVB group (Ge, Sn, Pb) in the valence bands to repel the CBM especially the Cd-*s* states up as the *s* states are rather delocalized. For the IVB-IIA-IIA-IIB group having IIB and IVB elements (e.g. $Cs_4$[PbSrBaCd]$I_{12}$), the amount of IVB elements in the alloy is not enough to push the IIB-*s* states up to induce upward band gap bowing. For the IIA-IIA-IIA-IIB group (e.g. $Cs_4$[CaSrBaCd]$I_{12}$) or for the IVB-IVB-IVB-IIA, IVB-IVB-IIA-IIA, and IVB-IIA-IIA-IIA groups (e.g. $Cs_4$[GeSnPbBa]$I_{12}$), either the IVB elements are missing or the IIB elements are missing in the alloys. In the IVB-IVB-IVB-IIA, IVB-IVB-IIA-IIA, or IVB-IIA-IIA-IIA group alloys, the *s* states of IIA are found to be not at the CBM (see Tables V and VI in Appendix C), thus lacking the repulsion between the IVB-*s* and IIA-*s* band-edge states for potential upward band gap bowing. In these materials, the inside-conduction-band or inside-valence-band state repulsion leads to downward band gap bowing.

Analogously to the inside-valence band level repulsion and the inside-conduction band level repulsion which lead to the downward band gap bowing [1-15], we propose a valence-to conduction band repulsion mechanism for the upward band gap bowing. This type of cross-band-gap *s*-*s* band repulsion coincides with the *s*-*s* repulsion as predicted in Bi-doped $In_2O_3$ [46] and $Ga_2O_3$ [47]. If the IIB-*s* states are pushed up far away from the CBM in the $A_4$[BB′B′′B′′′]$X_{12}$ alloys, the upward band gap bowing will be lost, and the IVB-s coupling within the valence band and the IVB-*p* coupling within the conduction band will lead to downward band gap bowing. To illustrate the mechanism,consider the alloy example $Cs_4$[SnPbSrCd]$I_{12}$ which has an upward band bowing with a small the excess band gap $\Delta E_g$ of 0.04 eV Here we artificially tune the original upward band gap bowing to downward bowing by adding an on-site U potential exclusively on Cd-s orbitals for both the alloy $Cs_4$[SnPbSrCd]$I_{12}$ and the pure compound $CsCdI_3$. As the on-site U increases, both the band gaps of the alloy $Cs_4$[SnPbSrCd]$I_{12}$ and compound $CsCdI_3$ goes up. The on-site U term pushes the unoccupied Cd-s orbital up, resulting in the transition from upward bowing to downward bowing at approximately U = 4 eV, as shown in Fig. 10(a). We show two snapshots as for a direct visualization: upward bowing at U = 0 eV vs downward bowing at U = 15 eV [Fig. 10(b,c)].

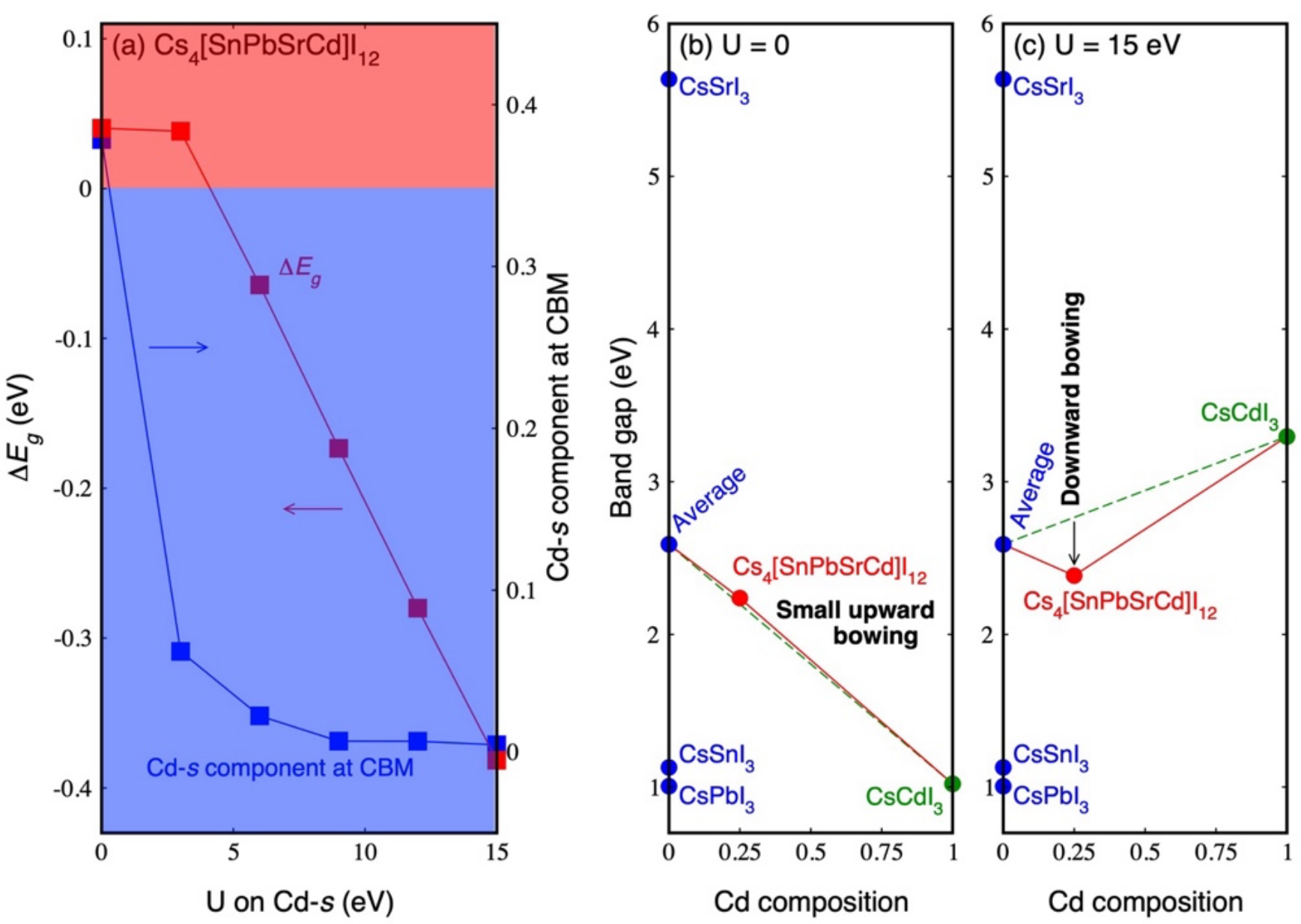

FIG. 10. Band gap bowing effects in $Cs_4[SnPbSrCd]I_{12}$. (a) $\Delta E_g$ of $Cs_4[SnPbSrCd]I_{12}$ as a function of U on Cd-states. Red curve: $\Delta E_g$. Blue curve: Cd-*s* component at CBM. (b,c) Band gap of $Cs_4[SnPbSrCd]I_{12}$ relative to the components at U = 0 and U = 15 eV, respectively.

In the above discussed cross-band-gap *s*-*s* band repulsion mechanism of upward band gap bowing between IIB-*s* and IVB-*s* states, the IIB-*s* states in the conduction bands repel the IVB-*s* states in the valence bands, pushing the valence bands down, increasing the material stability of the alloys as the total energy has the contribution from the sum of the occupied eigenvalues. Therefore, the cross-band-gap *s*-*s* band repulsion mechanism for upward band gap bowing also tends to stabilize the alloy. Whereas in $A_4$[BB′B′′B′′′]$X_{12}$ alloys with downward band gap bowing, e.g. $Cs_4[GeSnPbCa]I_{12}$ ($\Delta E_g^{alloy}$ = -0.23 eV), the Pb-*s* (or Ge-*s*) states in the valence bands repel the Sn-*s* states in the valence bands, pushing the upper valence bands up, decreasing the material stability of the alloys. There are other effects that can lead to upward band gap bowing in the $A_4$[BB′B′′B′′′]$X_{12}$ halide perovskite alloys, such as the volume effect and the SOC effect [18], as well as the state-non-coupling effect [16]—these effects do not necessarily lead to simultaneous upward band gap bowing and negative mixing enthalpy. Indeed, some of the two-component halide perovskite alloys (see Fig. 8) from Ref. [18] have upward band gap bowing but without Cd or other group IIB elements and missing the cross-band-gap *s*-*s* band repulsion mechanism—the upward band gap bowing in these alloys is thus not induced by the cross-band-gap *s*-*s* band repulsion but by other effects [16,18] and is thus not correlated with negative mixing enthalpies. The indicative relationship between the target functionality (here upward band gap bowing) and material stability as found in the four-component $A_4$[BB′B′′B′′′]$X_{12}$ halide perovskite alloys is sharply contrast to contraindicative relationship between the topological insulation and material stability as found in $ABO_3$ compounds [19]. The indicative relationship between the target functionality and material stability due to unique mechanisms, here, the B-site *s*-*s* repulsion in perovskites, could facilitate highly efficient design of functional materials.

## IV. SUMMARY

We find an interesting trend of upward band gap bowing effects in the multi-component (N = 4) B-site halide perovskite alloys $A_4[BB'B''B''']X_{12}$, suggesting a new mechanism for upward band gap bowing effects in bulk alloys, here the B-site *s*-*s* repulsion in halide perovskite alloys. Such cross-band-gap B-site IIB-*s* IVB-*s* band repulsion mechanism of upward band gap bowing pushes the valence IVB-*s* bands down, increasing the material stability of the alloys. Analogous trends also appear in the corresponding three-component and two-component halide perovskite alloys. We find that one of the predicted halide perovskite alloys, $Cs_4[GeSnPbCd]I_{12}$ has a band gap much larger than all its components. This type of upward-band-gap-bowing effect can be used to construct a different type of semiconductor interface that consists of two narrow-gap semiconductors and a wide-gap barrier layer or tunneling layer between them formed by the alloy due to the interface atomic interchange between the two narrow-gap semiconductors. $Cs_4[GeSnPbCd]I_{12}$ is also found to have a negative coherent mixing enthalpy relative to the polymorphous structures of cubic single perovskites. Thus, one can satisfy the simultaneous conditions of low mixing enthalpy and upward band gap bowing in the halide perovskite alloys. The interband *s*-*s* repulsion mechanism for upward band gap bowing in the calculated halide perovskite alloys suggest a new approach for designing functional materials, i.e. identifying mechanisms and the associated electronic structures that can enable the target material functionality and the material stability simultaneously.

### APPENDIX A: COMPUTATIONAL METHODS

To evaluate the stability and electronic structures of halide perovskite materials, we apply the generalized gradient approximation (GGA) to density functional theory (DFT) [48] using the projector-augmented wave (PAW) pseudopotential [49] (i.e. Cs_sv, Br, I, Ge_d, Sn_d, Pb_d, Cd, Ca_sv, Sr_sv, and Ba_sv) with the Perdew-Burke-Ernzerhof exchange-correlation functional [50] as implemented in the Vienna Ab initio Simulation Package (VASP) [51,52], with the PBEsol exchange correlation functional [53]. We use the plane wave basis set energy-cutoff of 405 eV, and the following reciprocal space K grids for the ordered structures as observed in experiments for the $ABX_3$ compounds with A = Cs, X = Br or I, and B = Ge, Sn, Pb, Ca, Sr, Ba, or Cd [26,29-34]: cubic perovskite (SG: Pm-3m) [12 × 12 × 12], orthorhombic perovskite (SG: Pnma) [8 × 8 × 4], hexagonal $CsCdBr_3$-type (SG: $P6_3/mmc$) [$8 \times 8 \times 8$], rhombohedral $CsGeBr_3$-type (SG: R3m) [$12 \times 12 \times 12$], orthorhombic $CsPbBr_3$-type (SG: Pnma) [$6 \times 12 \times 4$], tetragonal $CsSnI_3$ (SG: P4/mbm) [$6 \times 6 \times 10$], and orthorhombic $CsSrI_3$-type (SG: Cmcm) [$16 \times 4 \times 4$].

We calculate the band gaps of the materials by using the HSE06 hybrid functional [54] with 43% of exact exchange, which can describe very well the band gaps of $CsPbBr_3$ and $CdPbI_3$ [18], and spin-orbit coupling, which is taken into account by a perturbation $\sum_{i,l,m} V_l^{SO} \mathbf{L} \cdot \mathbf{S} |l,m\rangle_{ii} \langle l,m|$ to the pseudopotential, where $|l,m\rangle_i$ is the angular momentum eigenstate of the $i$th atomic site [55], for which we used the following reduced reciprocal space grids for the ordered structures as observed in experiments for the $ABX_3$ compounds [26,29-34]: cubic perovskite ($6 \times 6 \times 6$), orthorhombic perovskite with space group Pnma ($4 \times 4 \times 2$), hexagonal $CsCdBr_3$-type ($4 \times 4 \times 4$), rhombohedral $CsGeBr_3$-type ($6 \times 6 \times 6$), orthorhombic $CsPbBr_3$-type ($3 \times 6 \times 2$), tetragonal $CsSnI_3$ ($3 \times 3 \times 5$), and orthorhombic $CsSrI_3$-type ($8 \times 2 \times 2$), and the polymorphous $2\sqrt{2} \times 2\sqrt{2} \times 4$ supercell of cubic perovskite (Pm-3m) structure ($1 \times 1 \times 1$).

## APPENDIX B: DFT DATA ON BAND GAP BOWING AND MATERIAL STABILITY RELATIVE TO POLYMORPHOUS AND MONOMORPHOUS CUBIC $ABX_3$ CONSTITUENTS

Tables I and II (III and IV) list the coherent mixing enthalpies and the excess band gaps relative to polymorphous (monomorphous) cubic $ABX_3$ constituents of the four-component halide perovskite alloys. We see from Tables III and IV that by using the monomorphous cubic structures for the constituents (ideally ordered and with a single local environment), the alloy stability and the upward band gap bowing are significantly overestimated as compared to the results when using the polymorphous cubic structures (Table I and II) for the constituents. Especially, the calculated iodide alloys have negative mixing enthalpies relative to the monomorphous cubic constituents $ABX_3$ because the monomorphous constituents have abnormally high energy, not benefitting from polymorphous relaxation (see also Fig. 9 in the main text).

**Table I.** *Bromide polymorphous* band gaps, excess band gaps [Eq. (1)], coherent and incoherent mixing enthalpies of the four-component halide perovskite alloys $A_4[BB'B''B''']X_{12}$, where A = Cs; X = Br; B, B', B'', and B''' are four distinguished elements chosen from Group IVB Ge, Sn, Pb; Group IIB Cd; and Group IIA Ca, Sr, and Ba. Here the excess band gap of alloys refers to the band gap relative to the linear interpolation of polymorphous band gaps of pure compounds.

| Group of mixing B-site elements | Formula of B-site mixing alloy x= 25% each site | Band gap $E_g$ (eV) | Excess band gap $\Delta E_g$ (eV) | $\Delta H_{coh.}$ (meV/atom) | $\Delta H_{inc.}$ (meV/atom) | Number of alloy types |
|---|---|---|---|---|---|---|
| IVB-IVB-IVB-IIB | $Cs_4[GeSnPbCd]Br_{12}$ | 2.413 | 0.483 | -1.9 | 12.2 | 1 |
| | $Cs_4[GePbCaCd]Br_{12}$ | 3.184 | 0.031 | 2.5 | 16.6 | |
| | $Cs_4[GePbSrCd]Br_{12}$ | 3.254 | 0.070 | 4.9 | 17.9 | |
| | $Cs_4[GePbBaCd]Br_{12}$ | 3.332 | 0.238 | 5.6 | 28.0 | |
| | $Cs_4[GeSnCaCd]Br_{12}$ | 2.688 | -0.486 | 4.9 | 14.3 | |
| IVB-IVB-IIA-IIB | $Cs_4[GeSnSrCd]Br_{12}$ | 2.700 | -0.506 | 7.6 | 15.9 | 9 |
| | $Cs_4[GeSnBaCd]Br_{12}$ | 2.774 | -0.341 | 9.3 | 27.0 | |
| | $Cs_4[SnPbCaCd]Br_{12}$ | 2.918 | -0.192 | -2.1 | 12.7 | |
| | $Cs_4[SnPbSrCd]Br_{12}$ | 2.883 | -0.259 | -0.5 | 13.2 | |
| | $Cs_4[SnPbBaCd]Br_{12}$ | 2.961 | -0.090 | -0.7 | 22.4 | |
| | $Cs_4[GeCaBaCd]Br_{12}$ | 3.303 | -1.035 | 11.8 | 29.4 | |
| | $Cs_4[GeCaSrCd]Br_{12}$ | 3.271 | -1.158 | 9.0 | 17.2 | |
| | $Cs_4[GeSrBaCd]Br_{12}$ | 3.452 | -0.918 | 13.9 | 30.5 | |
| | $Cs_4[SnCaBaCd]Br_{12}$ | 2.988 | -1.307 | 5.6 | 23.9 | |
| IVB-IIA-IIA-IIB | $Cs_4[SnCaSrCd]Br_{12}$ | 3.043 | -1.344 | 4.3 | 13.2 | 9 |
| | $Cs_4[SnSrBaCd]Br_{12}$ | 3.021 | -1.306 | 5.8 | 23.0 | |
| | $Cs_4[PbCaBaCd]Br_{12}$ | 3.771 | -0.503 | 1.2 | 24.2 | |
| | $Cs_4[PbCaSrCd]Br_{12}$ | 3.839 | -0.526 | -0.7 | 12.9 | |
| | $Cs_4[PbSrBaCd]Br_{12}$ | 3.789 | -0.516 | 0.2 | 22.1 | |
| IIA-IIA-IIA-IIB | $Cs_4[CaSrBaCd]Br_{12}$ | 5.058 | -0.492 | 5.9 | 23.0 | 1 |
| | $Cs_4[GeSnPbCa]Br_{12}$ | 2.436 | -0.431 | 3.2 | 11.7 | |
| IVB-IVB-IVB-IIA | $Cs_4[GeSnPbSr]Br_{12}$ | 2.462 | -0.437 | 5.3 | 12.7 | 3 |
| | $Cs_4[GeSnPbBa]Br_{12}$ | 2.480 | -0.327 | 7.5 | 24.4 | |
| | $Cs_4[GePbCaBa]Br_{12}$ | 3.503 | -0.527 | 9.5 | 26.3 | |
| | $Cs_4[GePbCaSr]Br_{12}$ | 3.487 | -0.634 | 6.1 | 13.4 | |
| | $Cs_4[GePbSrBa]Br_{12}$ | 3.563 | -0.499 | 9.5 | 25.2 | |
| | $Cs_4[GeSnCaBa]Br_{12}$ | 3.231 | -0.821 | 12.8 | 25.0 | |
| IVB-IVB-IIA-IIA | $Cs_4[GeSnCaSr]Br_{12}$ | 3.235 | -0.907 | 10.9 | 13.6 | 9 |
| | $Cs_4[GeSnSrBa]Br_{12}$ | 3.197 | -0.886 | 14.2 | 25.3 | |
| | $Cs_4[SnPbCaBa]Br_{12}$ | 3.113 | -0.875 | -0.3 | 17.2 | |
| | $Cs_4[SnPbCaSr]Br_{12}$ | 3.082 | -0.996 | -0.9 | 7.2 | |
| | $Cs_4[SnPbSrBa]Br_{12}$ | 3.041 | -0.978 | 2.0 | 18.3 | |
| | $Cs_4[GeCaSrBa]Br_{12}$ | 4.178 | -1.128 | 15.1 | 26.1 | |
| IVB-IIA-IIA-IIA | $Cs_4[SnCaSrBa]Br_{12}$ | 3.771 | -1.492 | 8.3 | 19.9 | 3 |
| | $Cs_4[PbCaSrBa]Br_{12}$ | 4.224 | -1.018 | 1.5 | 17.8 | |

**Table II.** *Iodide polymorphous* band gaps, excess band gaps $\Delta E_g$ [Eq. (1)], coherent and incoherent mixing enthalpies of the four-component halide perovskite alloys $A_4[BB'B''B''']X_{12}$, where A = Cs; X = I; B, B′, B′′, and B′′′ are four distinguished elements chosen from Group IVB Ge, Sn, Pb; Group IIB Cd; and Group IIA Ca, Sr, and Ba. Here the excess band gap of alloys refers to the band gap relative to the linear interpolation of band gaps of pure polymorphous compounds.

| Group of mixing B-site elements | Formula of B-site mixing alloy x= 25% each site | Band gap $E_g$ (eV) | Excess band gap $\Delta E_g$ (eV) | $\Delta H_{coh.}$ (meV/atom) | $\Delta H_{inc.}$ (meV/atom) | Number of alloy types |
|---|---|---|---|---|---|---|
| **IVB-IVB-IVB-IIB** | $Cs_4[GeSnPbCd]I_{12}$ | 1.963 | 0.793 | -6.3 | 16.3 | 1 |
| **IVB-IVB-IIA-IIB** | $Cs_4[GePbCaCd]I_{12}$ | 2.617 | 0.388 | -6.2 | 22.8 | 9 |
| | $Cs_4[GePbSrCd]I_{12}$ | 2.704 | 0.406 | -0.6 | 24.0 | |
| | $Cs_4[GePbBaCd]I_{12}$ | 2.609 | 0.382 | -1.5 | 30.3 | |
| | $Cs_4[GeSnCaCd]I_{12}$ | 2.341 | 0.081 | -0.2 | 18.6 | |
| | $Cs_4[GeSnSrCd]I_{12}$ | 2.442 | 0.113 | 6.1 | 20.5 | |
| | $Cs_4[GeSnBaCd]I_{12}$ | 2.391 | 0.133 | 5.2 | 26.8 | |
| | $Cs_4[SnPbCaCd]I_{12}$ | 2.354 | 0.225 | -9.1 | 18.4 | |
| | $Cs_4[SnPbSrCd]I_{12}$ | 2.238 | 0.040 | -1.8 | 21.3 | |
| | $Cs_4[SnPbBaCd]I_{12}$ | 2.242 | 0.114 | -1.9 | 28.5 | |
| **IVB-IIA-IIA-IIB** | $Cs_4[GeCaBaCd]I_{12}$ | 3.063 | -0.254 | 5.4 | 33.4 | 9 |
| | $Cs_4[GeCaSrCd]I_{12}$ | 2.970 | -0.417 | 3.5 | 24.3 | |
| | $Cs_4[GeSrBaCd]I_{12}$ | 3.080 | -0.306 | 8.8 | 32.4 | |
| | $Cs_4[SnCaBaCd]I_{12}$ | 2.460 | -0.757 | 1.3 | 27.9 | |
| | $Cs_4[SnCaSrCd]I_{12}$ | 2.481 | -0.807 | -0.2 | 19.1 | |
| | $Cs_4[SnSrBaCd]I_{12}$ | 2.787 | -0.499 | 7.2 | 29.3 | |
| | $Cs_4[PbCaBaCd]I_{12}$ | 3.176 | -0.010 | -6.9 | 29.8 | |
| | $Cs_4[PbCaSrCd]I_{12}$ | 3.041 | -0.215 | -8.1 | 21.4 | |
| | $Cs_4[PbSrBaCd]I_{12}$ | 3.059 | -0.196 | -4.3 | 28.0 | |
| **IIA-IIA-IIA-IIB** | $Cs_4[CaSrBaCd]I_{12}$ | 3.860 | -0.485 | 1.1 | 29.6 | 1 |
| **IVB-IVB-IVB-IIA** | $Cs_4[GeSnPbCa]I_{12}$ | 2.026 | -0.231 | -1.6 | 19.2 | 3 |
| | $Cs_4[GeSnPbSr]I_{12}$ | 2.022 | -0.303 | 4.2 | 20.6 | |
| | $Cs_4[GeSnPbBa]I_{12}$ | 2.095 | -0.159 | 8.2 | 31.8 | |
| **IVB-IVB-IIA-IIA** | $Cs_4[GePbCaBa]I_{12}$ | 2.836 | -0.477 | 4.5 | 34.4 | 9 |
| | $Cs_4[GePbCaSr]I_{12}$ | 2.936 | -0.447 | -0.8 | 21.9 | |
| | $Cs_4[GePbSrBa]I_{12}$ | 2.971 | -0.410 | 3.8 | 29.4 | |
| | $Cs_4[GeSnCaBa]I_{12}$ | 2.709 | -0.635 | 8.3 | 28.1 | |
| | $Cs_4[GeSnCaSr]I_{12}$ | 2.704 | -0.710 | 7.2 | 19.7 | |
| | $Cs_4[GeSnSrBa]I_{12}$ | 2.770 | -0.643 | 12.5 | 27.8 | |
| | $Cs_4[SnPbCaBa]I_{12}$ | 2.559 | -0.654 | -5.2 | 23.3 | |
| | $Cs_4[SnPbCaSr]I_{12}$ | 2.563 | -0.721 | -5.6 | 15.7 | |
| | $Cs_4[SnPbSrBa]I_{12}$ | 2.528 | -0.753 | 3.5 | 27.6 | |
| **IVB-IIA-IIA-IIA** | $Cs_4[GeCaSrBa]I_{12}$ | 3.474 | -0.997 | 8.9 | 30.5 | 3 |
| | $Cs_4[SnCaSrBa]I_{12}$ | 3.208 | -1.163 | 4.5 | 24.7 | |
| | $Cs_4[PbCaSrBa]I_{12}$ | 3.469 | -0.871 | -5.1 | 25.4 | |

**Table III.** *Bromide monomorphous* excess band gaps relative to the linear interpolation of band gaps of cubic $ABX_3$ ($\Delta E_g^{mono.}$), (coherent) mixing enthalpies relative to the monomorphous cubic $ABX_3$ of the four-component halide perovskite alloys $A_4[BB'B''B''']X_{12}$, where A = Cs; X = Br; B, B′, B′′, and B′′′ are four distinguished elements chosen from Group IVB Ge, Sn, Pb; Group IIB Cd; and Group IIA Ca, Sr, and Ba. Here the excess band gap of alloys refers to the band gap relative to the linear interpolation of band gaps of pure monomorphous compounds.

| Group of mixing B-site elements | Formula of B-site mixing alloy x= 25% each site | $\Delta E_g^{mono.}$ (eV) | $\Delta H_{coh.}^{mono.}$ (meV/atom) |
|---|---|---|---|
| IVB-IVB-IVB-IIB | $Cs_4[GeSnPbCd]Br_{12}$ | 0.834 | -2.3 |
| IVB-IVB-IIA-IIB | $Cs_4[GePbCaCd]Br_{12}$ | 0.172 | 2.5 |
| | $Cs_4[GePbSrCd]Br_{12}$ | 0.317 | -1.7 |
| | $Cs_4[GePbBaCd]Br_{12}$ | 0.461 | -8.7 |
| | $Cs_4[GeSnCaCd]Br_{12}$ | -0.126 | 4.5 |
| | $Cs_4[GeSnSrCd]Br_{12}$ | -0.039 | 0.5 |
| | $Cs_4[GeSnBaCd]Br_{12}$ | 0.101 | -5.6 |
| | $Cs_4[SnPbCaCd]Br_{12}$ | 0.003 | -2.6 |
| | $Cs_4[SnPbSrCd]Br_{12}$ | 0.043 | -7.7 |
| | $Cs_4[SnPbBaCd]Br_{12}$ | 0.186 | -15.6 |
| IVB-IIA-IIA-IIB | $Cs_4[GeCaBaCd]Br_{12}$ | -0.803 | -2.6 |
| | $Cs_4[GeCaSrCd]Br_{12}$ | -0.901 | 2.3 |
| | $Cs_4[GeSrBaCd]Br_{12}$ | -0.579 | -7.2 |
| | $Cs_4[SnCaBaCd]Br_{12}$ | -1.021 | -9.4 |
| | $Cs_4[SnCaSrCd]Br_{12}$ | -1.033 | -3.0 |
| | $Cs_4[SnSrBaCd]Br_{12}$ | -0.913 | -15.8 |
| | $Cs_4[PbCaBaCd]Br_{12}$ | -0.436 | -13.3 |
| | $Cs_4[PbCaSrCd]Br_{12}$ | -0.434 | -7.4 |
| | $Cs_4[PbSrBaCd]Br_{12}$ | -0.343 | -21.0 |
| IIA-IIA-IIA-IIB | $Cs_4[CaSrBaCd]Br_{12}$ | -0.309 | -15.3 |
| IVB-IVB-IVB-IIA | $Cs_4[GeSnPbCa]Br_{12}$ | -0.079 | 2.8 |
| | $Cs_4[GeSnPbSr]Br_{12}$ | 0.022 | -1.8 |
| | $Cs_4[GeSnPbBa]Br_{12}$ | 0.107 | -7.3 |
| IVB-IVB-IIA-IIA | $Cs_4[GePbCaBa]Br_{12}$ | -0.303 | -4.9 |
| | $Cs_4[GePbCaSr]Br_{12}$ | -0.385 | -0.6 |
| | $Cs_4[GePbSrBa]Br_{12}$ | -0.168 | -11.5 |
| | $Cs_4[GeSnCaBa]Br_{12}$ | -0.377 | -2.1 |
| | $Cs_4[GeSnCaSr]Br_{12}$ | -0.439 | 3.8 |
| | $Cs_4[GeSnSrBa]Br_{12}$ | -0.336 | -7.3 |
| | $Cs_4[SnPbCaBa]Br_{12}$ | -0.596 | -15.3 |
| | $Cs_4[SnPbCaSr]Br_{12}$ | -0.693 | -8.1 |
| | $Cs_4[SnPbSrBa]Br_{12}$ | -0.593 | -19.7 |
| IVB-IIA-IIA-IIA | $Cs_4[GeCaSrBa]Br_{12}$ | -0.788 | -6.0 |
| | $Cs_4[SnCaSrBa]Br_{12}$ | -1.098 | -13.4 |
| | $Cs_4[PbCaSrBa]Br_{12}$ | -0.843 | -19.7 |

**Table IV.** Iodide *monomorphous* excess band gaps relative to the linear interpolation of band gaps of monomorphous cubic $ABX_3$ ($\Delta E_g^{mono.}$), (coherent) mixing enthalpies relative to the monomorphous cubic $ABX_3$ of the four-component halide perovskite alloys $A_4[BB'B''B''']X_{12}$, where A = Cs; X = I; B, B', B'', and B''' are four distinguished elements chosen from Group IVB Ge, Sn, Pb; Group IIB Cd; and Group IIA Ca, Sr, and Ba. Here the excess band gap of alloys refers to the band gap relative to the linear interpolation of band gaps of pure *monomorphous* compounds.

| **Group of mixing B-site elements** | **Formula of B-site mixing alloy x= 25% each site** | $\Delta E_g^{mono.}$ **(eV)** | $\Delta H_{coh.}^{mono.}$ **(meV/atom)** |
|---|---|---|---|
| **IVB-IVB-IVB-IIB** | $Cs_4[GeSnPbCd]I_{12}$ | 1.108 | -8.2 |
| **IVB-IVB-IIA-IIB** | $Cs_4[GePbCaCd]I_{12}$ | 0.530 | -6.2 |
| | $Cs_4[GePbSrCd]I_{12}$ | 0.663 | -9.6 |
| | $Cs_4[GePbBaCd]I_{12}$ | 0.610 | -16.8 |
| | $Cs_4[GeSnCaCd]I_{12}$ | 0.410 | -2.1 |
| | $Cs_4[GeSnSrCd]I_{12}$ | 0.557 | -5.5 |
| | $Cs_4[GeSnBaCd]I_{12}$ | 0.547 | -12.7 |
| | $Cs_4[SnPbCaCd]I_{12}$ | 0.377 | -11.7 |
| | $Cs_4[SnPbSrCd]I_{12}$ | 0.308 | -14.1 |
| | $Cs_4[SnPbBaCd]I_{12}$ | 0.353 | -20.5 |
| **IVB-IIA-IIA-IIB** | $Cs_4[GeCaBaCd]I_{12}$ | -0.012 | -9.9 |
| | $Cs_4[GeCaSrCd]I_{12}$ | -0.146 | -5.5 |
| | $Cs_4[GeSrBaCd]I_{12}$ | 0.051 | -16.3 |
| | $Cs_4[SnCaBaCd]I_{12}$ | -0.505 | -17.3 |
| | $Cs_4[SnCaSrCd]I_{12}$ | -0.526 | -12.6 |
| | $Cs_4[SnSrBaCd]I_{12}$ | -0.132 | -21.1 |
| | $Cs_4[PbCaBaCd]I_{12}$ | 0.055 | -23.0 |
| | $Cs_4[PbCaSrCd]I_{12}$ | -0.121 | -17.8 |
| | $Cs_4[PbSrBaCd]I_{12}$ | -0.015 | -30.1 |
| **IIA-IIA-IIA-IIB** | $Cs_4[CaSrBaCd]I_{12}$ | -0.291 | -24.7 |
| **IVB-IVB-IVB-IIA** | $Cs_4[GeSnPbCa]I_{12}$ | 0.088 | -3.4 |
| | $Cs_4[GeSnPbSr]I_{12}$ | 0.131 | -7.3 |
| | $Cs_4[GeSnPbBa]I_{12}$ | 0.245 | -9.6 |
| **IVB-IVB-IIA-IIA** | $Cs_4[GePbCaBa]I_{12}$ | -0.246 | -10.8 |
| | $Cs_4[GePbCaSr]I_{12}$ | -0.187 | -9.7 |
| | $Cs_4[GePbSrBa]I_{12}$ | -0.064 | -21.1 |
| | $Cs_4[GeSnCaBa]I_{12}$ | -0.217 | -9.5 |
| | $Cs_4[GeSnCaSr]I_{12}$ | -0.263 | -4.4 |
| | $Cs_4[GeSnSrBa]I_{12}$ | -0.109 | -15.1 |
| | $Cs_4[SnPbCaBa]I_{12}$ | -0.412 | -23.8 |
| | $Cs_4[SnPbCaSr]I_{12}$ | -0.450 | -17.8 |
| | $Cs_4[SnPbSrBa]I_{12}$ | -0.396 | -24.8 |
| **IVB-IIA-IIA-IIA** | $Cs_4[GeCaSrBa]I_{12}$ | -0.637 | -16.1 |
| | $Cs_4[SnCaSrBa]I_{12}$ | -0.792 | -23.8 |
| | $Cs_4[PbCaSrBa]I_{12}$ | -0.687 | -30.8 |

## APPENDIX C: ATOMIC ORBITAL COMPONENTS OF THE BAND-EDGE STATES

Tables V and VI list the atomic orbital components of the band-edge electronic states of the four-component halide perovskite alloys $A_4[BB'B''B''']X_{12}$. We see that, for each alloy, there is always a combination of *s* and *p* states in the VBM or CBM, respectively, from the same species, e.g. Sn-*s* and Sn-*p* states in VBM and CBM, respectively, in the alloy $Cs_4[GeSnPbCd]I_{12}$. These same-species opposite-parity (e.g. *s* versus *p*) states usually have allowed optical transition between them if not forbidden by structural symmetries that are missing in realistic alloys.

**Table V.** *Bromide* atomic orbital components of the band-edge electronic states of the four-component halide perovskite alloys $A_4[BB'B''B''']X_{12}$, where A = Cs; X = Br; B, B′, B′′, and B′′′ are four distinguished elements chosen from Group IVB Ge, Sn, Pb; Group IIB Cd; and Group IIA Ca, Sr, and Ba.

| **Group of mixing B-site elements** | **Formula of B-site mixing alloy x= 25% each site** | **CBM atomic orbital components (%)** | **VBM atomic orbital components (%)** |
|---|---|---|---|
| **IVB-IVB-IVB-IIB** | $Cs_4[GeSnPbCd]Br_{12}$ | Pb-*p* (30), Ge-*p* (19), Cd-*s* (19) | Br-*p* (55), Sn-*s* (27), Ge-*s* (7) |
| **IVB-IVB-IIA-IIB** | $Cs_4[GePbCaCd]Br_{12}$ | Ge-*p* (32), Pb-*p* (31), Cd-*s* (14) | Br-*p* (62), Ge-*s* (29), Pb-*s* (5) |
| | $Cs_4[GePbSrCd]Br_{12}$ | Cd-*s* (35), Ge-*p* (27), Br-*p* (15) | Br-*p* (62), Ge-*s* (26), Pb-*s* (8) |
| | $Cs_4[GePbBaCd]Br_{12}$ | Pb-*p* (36), Ge-*p* (20), Cd-*s* (19) | Br-*p* (62), Ge-*s* (28), Pb-*s* (5) |
| | $Cs_4[GeSnCaCd]Br_{12}$ | Cd-*s* (36), Ge-*p* (28), Br-*p* (15) | Br-*p* (52), Sn-*s* (30), Ge-*s* (10) |
| | $Cs_4[GeSnSrCd]Br_{12}$ | Cd-*s* (36), Ge-*p* (28), Br-*p* (15) | Br-*p* (54), Sn-*s* (31), Ge-*s* (8) |
| | $Cs_4[GeSnBaCd]Br_{12}$ | Cd-*s* (37), Ge-*p* (26), Br-*p* (16) | Br-*p* (55), Sn-*s* (30), Ge-*s* (9) |
| | $Cs_4[SnPbCaCd]Br_{12}$ | Cd-*s* (34), Sn-*p* (28), Br-*p* (13) | Br-*p* (57), Sn-*s* (37), Pb-*s* (3) |
| | $Cs_4[SnPbSrCd]Br_{12}$ | Cd-*s* (33), Sn-*p* (30), Br-*p* (14) | Br-*p* (56), Sn-*s* (37), Pb-*s* (2) |
| | $Cs_4[SnPbBaCd]Br_{12}$ | Sn-*p* (31), Cd-*s* (31), Br-*p* (12) | Br-*p* (57), Sn-*s* (37), Pb-*s* (3) |
| **IVB-IIA-IIA-IIB** | $Cs_4[GeCaBaCd]Br_{12}$ | Cd-*s* (37), Ge-*p* (30), Br-*p* (14) | Br-*p* (63), Ge-*s* (32), Ge-*p* (2) |
| | $Cs_4[GeCaSrCd]Br_{12}$ | Cd-*s* (38), Ge-*p* (31), Br-*p* (13) | Br-*p* (63), Ge-*s* (32), Ge-*p* (2) |
| | $Cs_4[GeSrBaCd]Br_{12}$ | Cd-*s* (38), Ge-*p* (27), Br-*p* (16) | Br-*p* (64), Ge-*s* (31), Ge-*p* (2) |
| | $Cs_4[SnCaBaCd]Br_{12}$ | Cd-*s* (39), Sn-*p* (31), Br-*p* (13) | Br-*p* (56), Sn-*s* (40), Sn-*p* (2) |
| | $Cs_4[SnCaSrCd]Br_{12}$ | Cd-*s* (40), Sn-*p* (32), Br-*p* (12) | Br-*p* (55), Sn-*s* (40), Sn-*p* (2) |
| | $Cs_4[SnSrBaCd]Br_{12}$ | Cd-*s* (38), Sn-*p* (32), Br-*p* (13) | Br-*p* (57), Sn-*s* (40), Sn-*p* (1) |
| | $Cs_4[PbCaBaCd]Br_{12}$ | Pb-*p* (38), Cd-*s* (33), Br-*p* (14) | Br-*p* (69), Pb-*s* (28), Cd-*d* (1) |
| | $Cs_4[PbCaSrCd]Br_{12}$ | Pb-*p* (39), Cd-*s* (33), Br-*p* (14) | Br-*p* (68), Pb-*s* (29), Cd-*d* (1) |
| | $Cs_4[PbSrBaCd]Br_{12}$ | Pb-*p* (40), Cd-*s* (32), Br-*p* (14) | Br-*p* (70), Pb-*s* (29), Cd-*d* (1) |
| **IIA-IIA-IIA-IIB** | $Cs_4[CaSrBaCd]Br_{12}$ | Cd-*s* (58), Br-*p* (20), Br-*s* (12) | Br-*p* (99), Ba-*p* (1) |
| **IVB-IVB-IVB-IIA** | $Cs_4[GeSnPbCa]Br_{12}$ | Pb-*p* (49), Sn-*p* (18), Ge-*p* (17) | Br-*p* (56), Sn-*s* (28), Ge-*s* (8) |
| | $Cs_4[GeSnPbSr]Br_{12}$ | Pb-*p* (52), Ge-*p* (17), Sn-*p* (15) | Br-*p* (56), Sn-*s* (27), Ge-*s* (8) |
| | $Cs_4[GeSnPbBa]Br_{12}$ | Pb-*p* (54), Ge-*p* (15), Sn-*p* (13) | Br-*p* (56), Sn-*s* (28), Ge-*s* (7) |
| **IVB-IVB-IIA-IIA** | $Cs_4[GePbCaBa]Br_{12}$ | Pb-*p* (41), Ge-*p* (34), Br-*p* (12) | Br-*p* (62), Ge-*s* (25), Pb-*s* (10) |
| | $Cs_4[GePbCaSr]Br_{12}$ | Pb-*p* (40), Ge-*p* (36), Br-*p* (12) | Br-*p* (61), Ge-*s* (25), Pb-*s* (10) |
| | $Cs_4[GePbSrBa]Br_{12}$ | Pb-*p* (48), Ge-*p* (28), Br-*p* (12) | Br-*p* (62), Ge-*s* (24), Pb-*s* (11) |
| | $Cs_4[GeSnCaBa]Br_{12}$ | Ge-*p* (51), Sn-*p* (24), Br-*p* (9) | Br-*p* (56), Sn-*s* (31), Ge-*s* (10) |
| | $Cs_4[GeSnCaSr]Br_{12}$ | Ge-*p* (52), Sn-*p* (23), Br-*p* (9) | Br-*p* (55), Sn-*s* (31), Ge-*s* (10) |
| | $Cs_4[GeSnSrBa]Br_{12}$ | Ge-*p* (50), Sn-*p* (25), Br-*p* (9) | Br-*p* (55), Sn-*s* (33), Ge-*s* (9) |
| | $Cs_4[SnPbCaBa]Br_{12}$ | Pb-*p* (48), Sn-*p* (35), Br-*p* (9) | Br-*p* (57), Sn-*s* (36), Pb-*s* (5) |
| | $Cs_4[SnPbCaSr]Br_{12}$ | Pb-*p* (49), Sn-*p* (34), Br-*p* (8) | Br-*p* (56), Sn-*s* (37), Pb-*s* (4) |
| | $Cs_4[SnPbSrBa]Br_{12}$ | Pb-*p* (44), Sn-*p* (38), Br-*p* (9) | Br-*p* (56), Sn-*s* (37), Pb-*s* (4) |
| **IVB-IIA-IIA-IIA** | $Cs_4[GeCaSrBa]Br_{12}$ | Ge-*p* (71), Br-*p* (13), Br-s (7) | Br-*p* (65), Ge-*s* (31), Ge-*p* (2) |
| | $Cs_4[SnCaSrBa]Br_{12}$ | Sn-*p* (75), Br-*p* (13), Br-s (7) | Br-*p* (55), Sn-*s* (41), Sn-*p* (2) |
| | $Cs_4[PbCaSrBa]Br_{12}$ | Pb-*p* (72), Br-*p* (17), Br-s (7) | Br-*p* (67), Pb-*s* (31), Pb-*p* (1) |

**Table VI.** *Iodide* atomic orbital components of the band-edge electronic states of the four-component halide perovskite alloys $A_4[BB'B''B''']X_{12}$, where A = Cs; X = I; B, B′, B′′, and B′′′ are four distinguished elements chosen from Group IVB Ge, Sn, Pb; Group IIB Cd; and Group IIA Ca, Sr, and Ba.

| **Group of mixing B-site elements** | **Formula of B-site mixing alloy x= 25% each site** | **CBM atomic orbital components (%)** | **VBM atomic orbital components (%)** |
|---|---|---|---|
| **IVB-IVB-IVB-IIB** | $Cs_4[GeSnPbCd]I_{12}$ | Sn-*p* (17), Ge-*p* (14), Cd-*s* (12) | I-*p* (61), Sn-*s* (27), Pb-*s* (5) |
| **IVB-IVB-IIA-IIB** | $Cs_4[GePbCaCd]I_{12}$ | Pb-*p* (39), Ge-*p* (20), Cd-*s* (15) | I-*p* (68), Ge-*s* (24), Pb-*s* (4) |
| | $Cs_4[GePbSrCd]I_{12}$ | Pb-*p* (40), Ge-*p* (18), I-*p* (16), Cd-*s* (15) | I-*p* (68), Ge-*s* (23), Pb-*s* (4) |
| | $Cs_4[GePbBaCd]I_{12}$ | Pb-*p* (39), Ge-*p* (28), I-*p* (13), Cd-*s* (7) | I-*p* (68), Ge-*s* (24), Pb-*s* (4) |
| | $Cs_4[GeSnCaCd]I_{12}$ | Cd-*s* (39), I-*p* (21), Ge-*p* (20) | I-*p* (60), Sn-*s* (27), Ge-*s* (7) |
| | $Cs_4[GeSnSrCd]I_{12}$ | Cd-*s* (38), I-*p* (21), Ge-*p* (18) | I-*p* (62), Sn-*s* (29), Ge-*s* (5) |
| | $Cs_4[GeSnBaCd]I_{12}$ | Cd-*s* (36), Ge-*p* (19), I-*p* (19) | I-*p* (61), Sn-*s* (26), Ge-*s* (8) |
| | $Cs_4[SnPbCaCd]I_{12}$ | Cd-*s* (31), Sn-*p* (25), I-*p* (16) | I-*p* (62), Sn-*s* (33), Pb-*s* (3) |
| | $Cs_4[SnPbSrCd]I_{12}$ | Cd-*s* (38), Sn-*p* (24), I-*p* (18) | I-*p* (63), Sn-*s* (33), Pb-*s* (2) |
| | $Cs_4[SnPbBaCd]I_{12}$ | Cd-*s* (37), Sn-*p* (26), I-*p* (17) | I-*p* (62), Sn-*s* (33), Pb-*s* (2) |
| **IVB-IIA-IIA-IIB** | $Cs_4[GeCaBaCd]I_{12}$ | Cd-*s* (37), I-*p* (23), Ge-*p* (21) | I-*p* (69), Ge-*s* (27), Ge-*p* (3) |
| | $Cs_4[GeCaSrCd]I_{12}$ | Cd-*s* (40), Ge-*p* (21), I-*p* (21) | I-*p* (70), Ge-*s* (26), Ge-*p* (2) |
| | $Cs_4[GeSrBaCd]I_{12}$ | Cd-*s* (38), I-*p* (23), Ge-*p* (22) | I-*p* (70), Ge-*s* (26), Ge-*p* (3) |
| | $Cs_4[SnCaBaCd]I_{12}$ | Cd-*s* (39), Sn-*p* (27), I-*p* (18) | I-*p* (63), Sn-*s* (34), Sn-*p* (1) |
| | $Cs_4[SnCaSrCd]I_{12}$ | Cd-*s* (40), Sn-*p* (27), I-*p* (18) | I-*p* (62), Sn-*s* (34), Sn-*p* (2) |
| | $Cs_4[SnSrBaCd]I_{12}$ | Cd-*s* (36), Sn-*p* (28), I-*p* (20) | I-*p* (61), Sn-*s* (35), Sn-*p* (2) |
| | $Cs_4[PbCaBaCd]I_{12}$ | Pb-*p* (50), I-*p* (20), Cd-*s* (20) | I-*p* (78), Pb-*s* (20), Pb-*p* (1) |
| | $Cs_4[PbCaSrCd]I_{12}$ | Cd-*s* (35), Pb-*p* (31), I-*p* (20) | I-*p* (75), Pb-*s* (23), Cd-*d* (1) |
| | $Cs_4[PbSrBaCd]I_{12}$ | Cd-*s* (34), Pb-*p* (31), I-*p* (20) | I-*p* (76), Pb-*s* (22), Pb-*p* (1) |
| **IIA-IIA-IIA-IIB** | $Cs_4[CaSrBaCd]I_{12}$ | Cd-*s* (54), I-*p* (25), I-*s* (9) | I-*p* (98), Cd-*d* (1) |
| **IVB-IVB-IVB-IIA** | $Cs_4[GeSnPbCa]I_{12}$ | Pb-*p* (47), Ge-*p* (18), Sn-*p* (16) | I-*p* (59), Sn-*s* (27), Ge-*s* (7) |
| | $Cs_4[GeSnPbSr]I_{12}$ | Pb-*p* (50), Ge-*p* (18), Sn-*p* (13) | I-*p* (60), Sn-*s* (27), Ge-*s* (7) |
| | $Cs_4[GeSnPbBa]I_{12}$ | Pb-*p* (50), Ge-*p* (17), Sn-*p* (13) | I-*p* (60), Sn-*s* (26), Ge-*s* (6) |
| **IVB-IVB-IIA-IIA** | $Cs_4[GePbCaBa]I_{12}$ | Ge-*p* (40), Pb-*p* (32), I-*p* (14) | I-*p* (67), Ge-*s* (20), Pb-*s* (10) |
| | $Cs_4[GePbCaSr]I_{12}$ | Ge-*p* (38), Pb-*p* (35), I-*p* (14) | I-*p* (67), Ge-*s* (20), Pb-*s* (10) |
| | $Cs_4[GePbSrBa]I_{12}$ | Pb-*p* (40), Ge-*p* (34), I-*p* (14) | I-*p* (67), Ge-*s* (21), Pb-*s* (9) |
| | $Cs_4[GeSnCaBa]I_{12}$ | Ge-*p* (50), Sn-*p* (22), I-*p* (12) | I-*p* (60), Sn-*s* (29), Ge-*s* (8) |
| | $Cs_4[GeSnCaSr]I_{12}$ | Ge-*p* (51), Sn-*p* (21), I-*p* (12) | I-*p* (59), Sn-*s* (30), Ge-*s* (8) |
| | $Cs_4[GeSnSrBa]I_{12}$ | Ge-*p* (50), Sn-*p* (23), I-*p* (12) | I-*p* (60), Sn-*s* (30), Ge-*s* (7) |
| | $Cs_4[SnPbCaBa]I_{12}$ | Pb-*p* (41), Sn-*p* (39), I-*p* (10) | I-*p* (61), Sn-*s* (33), Pb-*s* (4) |
| | $Cs_4[SnPbCaSr]I_{12}$ | Pb-*p* (41), Sn-*p* (37), I-*p* (11) | I-*p* (62), Sn-*s* (34), Pb-*s* (4) |
| | $Cs_4[SnPbSrBa]I_{12}$ | Sn-*p* (47), Pb-*p* (33), I-*p* (9) | I-*p* (61), Sn-*s* (34), Pb-*s* (3) |
| **IVB-IIA-IIA-IIA** | $Cs_4[GeCaSrBa]I_{12}$ | Ge-*p* (68), I-*p* (16), I-*s* (7) | I-*p* (71), Ge-*s* (27), Ge-*p* (2) |
| | $Cs_4[SnCaSrBa]I_{12}$ | Sn-*p* (75), I-*p* (13), I-*s* (7) | I-*p* (61), Sn-*s* (35), Sn-*p* (2) |
| | $Cs_4[PbCaSrBa]I_{12}$ | Pb-*p* (69), I-*p* (19), I-*s* (7) | I-*p* (76), Pb-*s* (23), Pb-*p* (1) |

## ACKNOWLEDGEMENTS

This material is based upon work supported by the U.S. Department of Energy's Office of Energy Efficiency and Renewable Energy (EERE) under the Solar Energy Technologies Office Award No. DE-EE0009515. The research was performed using computational resources sponsored by the Department of Energy's Office of Energy Efficiency and Renewable Energy and located at the National Renewable Energy Laboratory.

## DATA AVAILABILITY

The data that support findings of this article are openly available [41].